\documentclass[11pt]{article}
\usepackage[margin=1in]{geometry}
\usepackage{lmodern}
\usepackage{sidecap}
\usepackage{xspace}
\usepackage{doi}
\usepackage{hyperref} % Should come last
\usepackage{tikz}
\usepackage{pgfgantt}
\usepackage{graphicx,pbox}
\usepackage{xcolor,hyperref,colortbl}
\usepackage{wrapfig}
\usepackage{enumitem}
\usepackage{gensymb}
\usepackage{comment}

\newcommand\pubnumber{SLAC-PUB-17660}
\newcommand\pubdate{\today}
\newcommand\pubblock{\rightline{\begin{tabular}{l} \pubnumber\\
          \pubdate \end{tabular}}}

\def\Title#1{\begin{center} {\Large #1 } \end{center}}
\def\Author#1{\begin{center}{ \sc #1} \end{center}}

\def\CCC{C$^{3}$~}
\def\CCCfive{C$^{3}$-550}
\def\CCCtwo{C$^{3}$-250}
\def\ee{e^+e^-}
\def\LN{LN$_2$~}

\newcommand\snowmass{\begin{center}\rule[-0.2in]{\hsize}{0.01in}\\\rule{\hsize}{0.01in}\\
\vskip 0.1in Submitted to the  Proceedings of the US Community Study\\ 
on the Future of Particle Physics (Snowmass 2021)\\ 
\rule{\hsize}{0.01in}\\\rule[+0.2in]{\hsize}{0.01in} \end{center}}
\begin{document}
\begin{titlepage}
\snowmass
\pubblock

%\vfill
\Title{\CCC Demonstration Research and Development Plan}

\Author{\textbf{Editors:}}

\Author{Emilio A. Nanni$^{7}$, Martin Breidenbach$^{7}$, Caterina Vernieri$^{7}$, Sergey Belomestnykh$^{3,8}$, Pushpalatha Bhat$^{3}$ and  Sergei Nagaitsev$^{3,11}$} 

\Author{\textbf{Authors:}} 

\Author{Mei Bai$^{7}$, Tim Barklow$^{7}$, William J. Berg$^{1}$, John Byrd$^{1}$, Ankur Dhar$^{7}$, Ram C. Dhuley$^{3}$, Chris Doss$^{10}$, Joseph Duris$^{7}$,  Auralee Edelen$^{7}$, Claudio Emma$^{7}$, Josef Frisch$^{7}$, Annika Gabriel$^{7}$, Spencer Gessner$^{7}$, Carsten Hast$^{7}$, Chunguang Jing$^{1}$, Arkadiy Klebaner$^{3}$, Anatoly K. Krasnykh$^{7}$, John Lewellen$^{7}$, Matthias Liepe$^{2}$, Michael Litos$^{10}$, Xueying Lu$^{1}$, Jared Maxson$^{2}$, David Montanari$^{3}$, Pietro Musumeci$^{9}$, Alireza Nassiri$^{1}$, Cho-Kuen Ng$^{7}$, Mohamed A. K. Othman$^{7}$, Marco Oriunno$^{7}$, Dennis Palmer$^{7}$,  J. Ritchie Patterson$^{2}$, Michael E. Peskin$^{7}$, Thomas J. Peterson$^{7}$, John Power$^{1}$, Ji Qiang$^{4}$, James Rosenzweig$^{9}$, Vladimir Shiltsev, Evgenya Simakov$^{5}$, Emma Snively$^{7}$, Bruno Spataro$^{6}$, Sami Tantawi$^{7}$, Brandon Weatherford$^{7}$, Glen White$^{7}$, and Kent P. Wootton$^{1}$}

\noindent
$^{1}${Argonne National Laboratory}\\
$^{2}${Cornell University}\\
$^{3}${Fermi National Accelerator Laboratory} \\
$^{4}${Lawrence Berkeley National Laboratory}\\
$^{5}${Los Alamos National Laboratory}\\
$^{6}${National Laboratory of Frascati, INFN-LNF}\\
$^{7}${SLAC National Accelerator Laboratory, Stanford University}\\
$^{8}${Stony Brook University}\\
$^{9}${University of California, Los Angeles} \\
$^{10}${University of Colorado, Boulder} \\
$^{11}${University of Chicago}\\

%\maketitle

%\date{February 2022}

%\maketitle
\end{titlepage}
\newpage

\tableofcontents
\newpage

\addcontentsline{toc}{section}{Executive Summary}
\section*{Executive Summary}
\CCC is an opportunity to realize an $\ee$ collider for the study of the Higgs boson at  $\sqrt{s} = 250$ GeV, with a well defined upgrade path to 550 GeV while staying on the same short facility footprint~\cite{c3phyiscswhitepaper,C3}. \CCC is based on a fundamentally new approach to normal conducting linear accelerators that achieves both high gradient and high efficiency at relatively low cost. Given the advanced state of linear collider designs, the key system that requires technical maturation for \CCC is the main linac. This white paper presents the staged approach towards a facility to demonstrate \CCC technology with both Direct (source and main linac) and Parallel (beam delivery, damping ring, ancillary component) R\&D. The white paper also includes discussion on the approach for technology industrialization, related HEP R\&D activities that are enabled by \CCC R\&D, infrastructure requirements and siting options. 

The primary goal of the \CCC Demonstration R\&D Plan is to reduce technical and cost risk by building and operating the key components of \CCC at an adequate scale. Each stage, see Table~\ref{tab:staging} has clear technical objective and deliverables that are synergistic with other accelerator based research fields. This R\&D plan starts with the engineering design, and demonstration of one cryomodule and will culminate in the construction of a 3 cryomodule linac with pre-production prototypes.  This R\&D program would also demonstrate the linac rf fundamentals including achievable gradient and gradient stability over a full electron bunch train and breakdown rates. It will also investigate beam dynamics including energy spread, wakefields, and emittance growth. This work will be critical to confirm the suitability of the \CCC beam parameters for the physics reach and detector performance in preparation for a Conceptual Design Report (CDR), as well as for follow-on technology development and industrialization. 

The \CCC Demonstration R\&D Plan will open up significant new scientific and technical opportunities based on development of high-gradient and high-efficiency accelerator technology. It will push this technology to operate both at the GeV scale and mature the technology to be reliable and provide high-brightness electron beams. 

The timeline for progressing with \CCC technology development will be governed by practical limitations on both the technical progress and resource availability. It consists of four stages: Stage~0) Ongoing fundamental R\&D on structure prototypes, damping and vibrations. Stage~1) Advancing the engineering maturity of the design and developing start-to-end simulations including space-charge and wakefield effects. Stage 1 will terminate with submission of a full proposal for the \CCC Demonstration R\&D Plan by the end of 2024. Stage~2) Production and testing of the first cryomodule at cryogenic temperatures. This would provide sufficient experimental data to compile a CDR and it is anticipated for Stage~2 to last 3 years (2025-2027) and to culminate with the transport of photo-electrons through the first cryomodule. Stage~3) Updates to the engineering design of the cryomodules, production of the second and third cryomodule and their installation. Beam tests would be performed with increasing beam current to test full beam loading. Lower charge and lower emittance beams will be used to investigate emittance growth. The successful full demonstration of the 3 cryomodules to deliver up to a 3 GeV beam and achieve the \CCCfive\ gradient will allow a comprehensive and robust evaluation of the technical design of \CCC as well as mitigate technical, schedule, and cost risks required to proceed with a Technical Design Report (TDR).

\begin{table}[ht!]
\begin{center}
\begin{tabular}{p{0.08\textwidth}p{0.08\textwidth}p{0.35\textwidth}p{0.4\textwidth}}%{c  c  c c  } 
\hline
 	& Time Frame &	Key R\&D	& Synergy and Spin-Offs \\
 \hline
 % & \CCC & \CCC \\
  % \hline\hline
  
Stage 0	& Ongoing &	Fundamental structure R\&D with prototype structure demonstration with beam and corresponding industrialization &	Cost effective compact linacs for medical, security and industrial applications (irradiation with electrons, x-rays) \\
Stage 1 & 2022-2024 & Beamline and cryogenics design study for demonstrator. Cryomodule engineering design and raft prototyping. & High brightness electron source and photo injector feasibility. Linacs for injection at scientific facility (injectors, booster, capture, \textit{etc.}) \\
Stage 2 & 2025-2027 & First high-gradient test with cryomodule. Implement one-cryomodule based linac to allow test with beam. &  \CCC based next generation X-FEL, beam dynamics study including beam loading, compact light sources  \\
Stage 3 & 2027-2029 & Develop the second and third cryomodules, demonstration with beam up to full beam loading. &  Future facility studies: Beam dynamics, positron targets, advanced concept based final focusing for linear collider, PWFA experiments, \textit{etc.}\\

\hline
 \end{tabular}
\end{center}
\caption{Staging of \CCC Demonstration R\&D Plan. A more detailed timeline is presented in Figure~\ref{fig:Timeline}}
 \label{tab:staging}
\end{table}

\newpage

\section{Introduction} 
Translating the high-gradients and exceptional rf efficiency of cryogenic-copper rf accelerator technology from laboratory prototypes to practical and scalable technology with GeV-class beams will have enormous benefits for HEP and many related accelerator applications. However, the technical and operational feasibility demonstration will require a dedicated and integrated R\&D plan that develops the modular units of the rf accelerator, control system, and rf sources, while addressing the key technical concerns for cryomodule engineering design, gradient, power, alignment, manufacturability, and industrialization. 

\CCC is an opportunity to realize an $\ee$ collider with this technology for the study of the Higgs boson at  $\sqrt{s} = 250$ GeV, with a well defined upgrade path to 550 GeV to study the top-Higgs Yukawa coupling and Higgs boson pair production, while staying on the same short facility footprint~\cite{c3phyiscswhitepaper}. The target beam parameters for \CCC are listed in Table~\ref{tab:LCparam}. \CCC can also be extended to TeV scale energies to probe new physics in unexplored/underexplored weakly coupled states that may well escape detection at the HL-LHC. 
\CCC is based on a fundamentally new approach to normal conducting linear accelerators that achieves both high gradient and high efficiency at relatively low cost. This design is described in~\cite{C3}. The primary goal of the \CCC Demonstration R\&D Plan is to reduce technical and cost risk by building and operating the key components of \CCC at an adequate scale. 

\begin{figure}[!ht]
\begin{center}
 \includegraphics[trim={0 0cm 0cm 0cm}, clip, width=0.95\hsize]{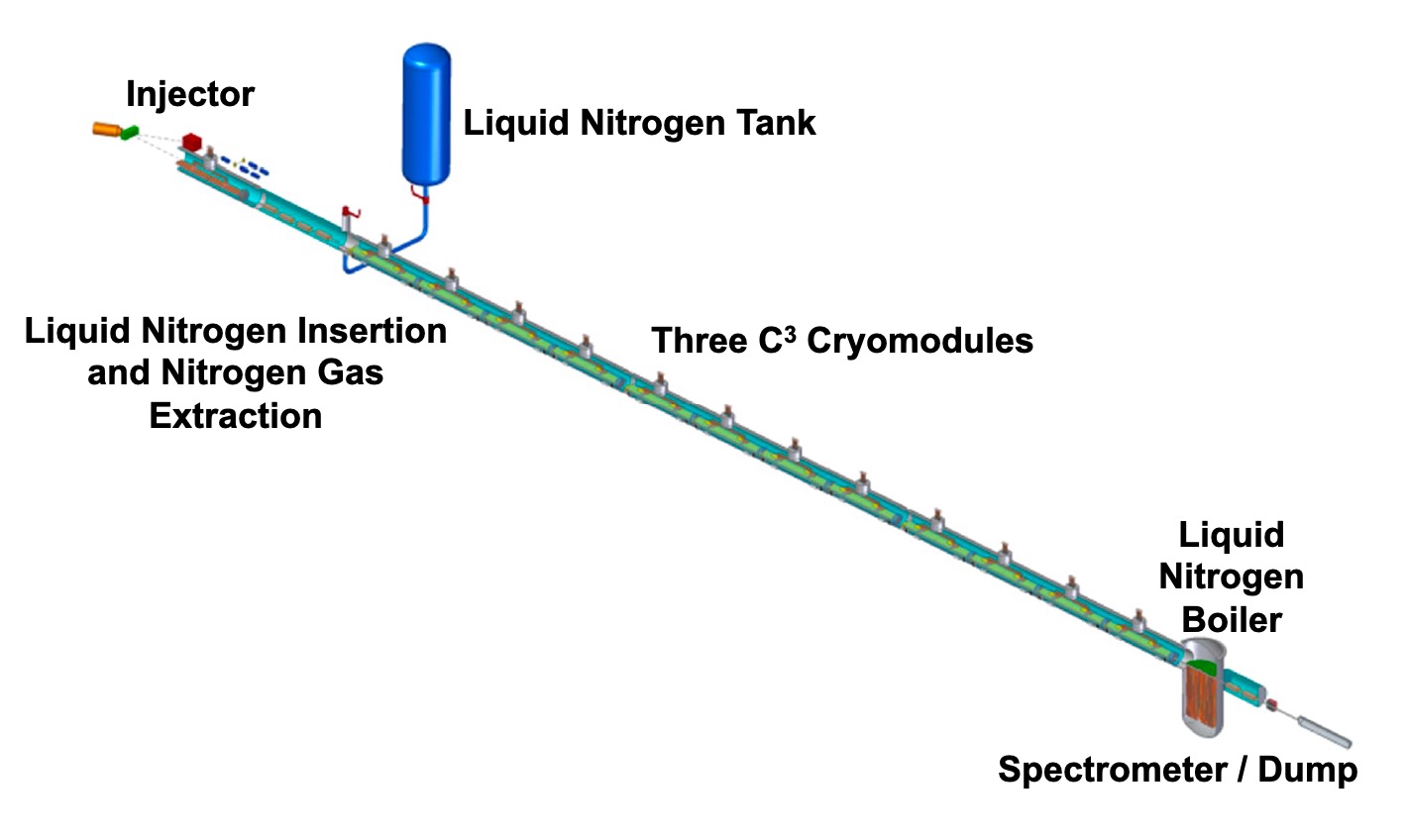} 
 \end{center}
\caption{Schematic of the full 3 cryomodule \CCC demonstrator capable of achieving all of the technical milestones for the R\&D plan. The full length of the layout it $\sim$50~m.}
\label{fig:layout}
\end{figure}

\begin{figure}[!ht]
%\begin{center}
 \includegraphics[trim={0 0cm 0cm 0cm}, clip, width=0.95\hsize]{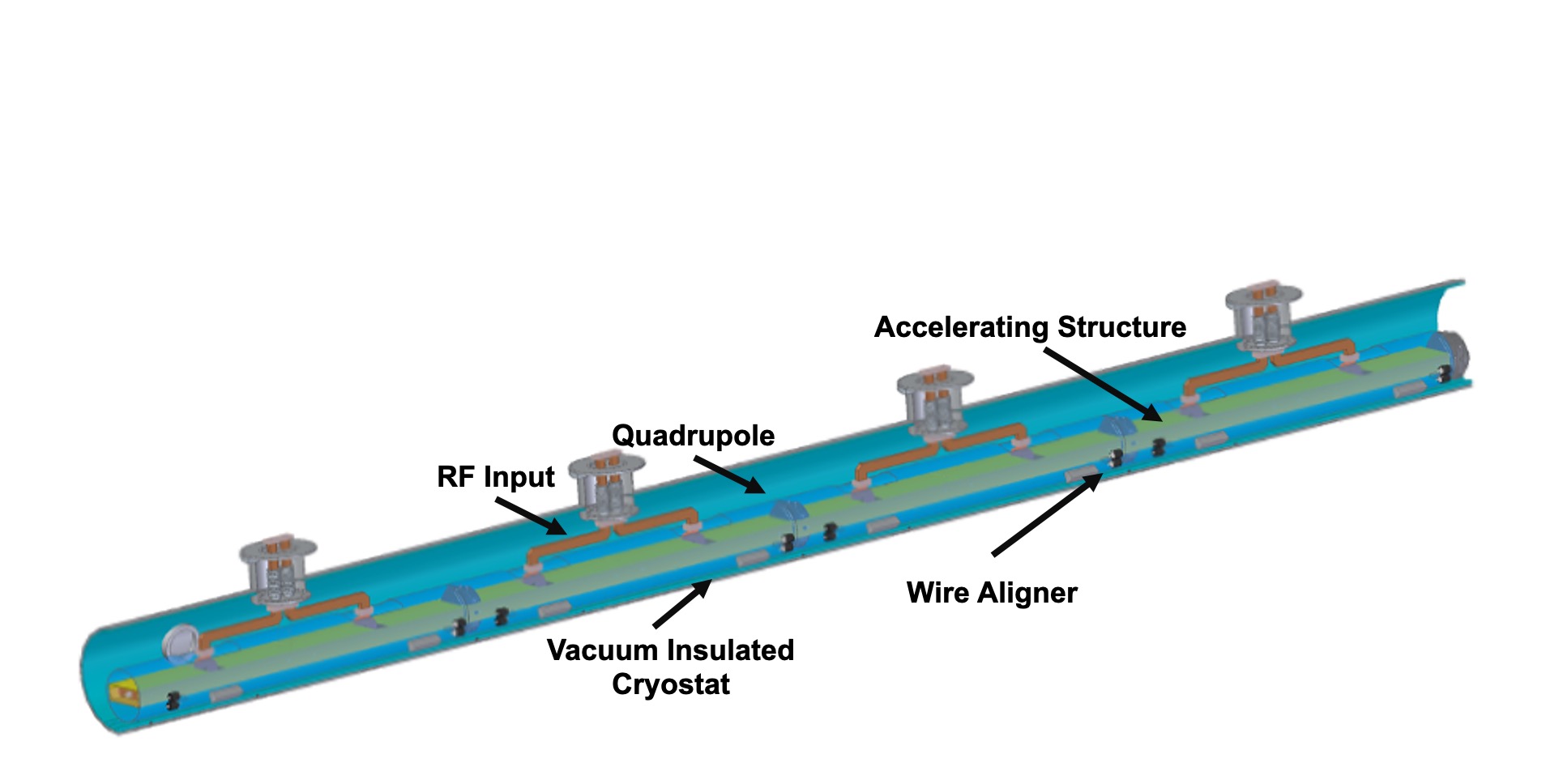} 
% \end{center}
\caption{Rendering of the 9~m \CCC cryomodule with 8 one meter accelerating structures, 4 two meter support rafts, 4 permanent magnetic quadrupoles and integrated beam position monitors.}
\label{fig:cryomod}
\end{figure}

Given the advanced state of linear collider designs, the key system that requires technical maturation for \CCC is the main linac. We propose a staged R\&D plan that will culminate in the construction of a 3 cryomodule linac with the cryomodules being pre-production prototypes. This linac would be fed by an S-band rf photo-injector, a booster linac and a magnetic bunch compressor. The injector would utilize accelerator technology that would be suitable for the electron source and positron capture linac.  The linac would feed an energy spectrometer for direct gradient measurements and then send the beam to a dump. This linac and electron source technology will serve as direct prototypes for FELs and other instruments requiring a high brightness, high gradient, compact linac. The full demonstration linac is shown in Fig.~\ref{fig:layout}. The target tolerances for the injector and cryomodule are listed in Table~\ref{tab:facility}. A more detailed view of a single cryomodule is shown in Figure~\ref{fig:cryomod}.

%Diagram of facility
The completion of the \CCC demonstration R\&D program would demonstrate:
\begin{itemize}
    \item 
the linac rf fundamentals including achievable gradient and gradient stability over a full electron bunch train and breakdown rates
\item beam dynamics including energy spread, wakefields, and emittance growth
\item cooling performance under liquid nitrogen (LN$_2$) with direct measurements of vibrations from several accelerators fitted with sensitive cryogenic accelerometers
\item small quantity component costs providing a reliable basis for extrapolation to \CCC scale production
\item prototype assembly tooling and exploration of industrial production and assembly
\item module-to-module assembly, possibly of simple robotic welding for cryostat connections
\end{itemize}

\begin{table}[ht!]
\begin{center}
\begin{tabular}{c  c  c } 
\hline
 Parameter [Unit] & Value & Value \\
 \hline
 % & \CCC & \CCC \\
  % \hline\hline
   CM Energy [GeV] & 250 & 550 \\
  Num. Bunches per Train  & 133 & 75 \\
  Train Rep. Rate [Hz] &  120 & 120 \\
  Bunch Spacing [ns] &  5.26 &  3.5 \\
  Bunch Charge [nC] & 1 & 1 \\
  Beam Power [MW] & 2 & 2.45\\
Gradient [MeV/m] & 70 & 120 \\
Effective Gradient [MeV/m] &  63 & 108 \\   %
Shunt Impedance [M$\Omega$/m] &  300 & 300 \\ %
Length [km]  & 8 & 8 \\

\hline
 \end{tabular}
\end{center}
\caption{Target beam parameters for \CCC.}
 \label{tab:LCparam}
\end{table}

The injector for the demonstrator linac will consist of an rf photo-injector with a removable CsTe cathode~\cite{PhysRevSTAB.18.043401} or another cathode material capable of producing suitable electron bunches. The goal will be to extract up to 133~nC of charge over 700~ns from the photo-injector to match the beam loading of the \CCC main linac. The individual bunch charge will vary depending on the beam test that is being conducted. Beam dynamics simulations of the injector and demonstrator linac will determine the target bunch charge and bunch spacing. The photo-injector will operate at S-band (a sub-harmonic of the C-band accelerator frequency). The photo-injector will produce 3-5~MeV electron bunches. This will be followed by matching optics to inject the beam into two one-meter long S-band accelerating structures operating at cryogenic temperatures. These linacs will operate at 50~MeV/m and will have a large transverse aperture to minimize any short- or long-range wakefields. The S-band section will bring the beam energy to 100~MeV. A magnetic chicane will be used to compress the electron bunches prior to injection into the first C-band cryomodule.

%%%%%%%%%%%%%%%%%%%%%%%%%%%%%%%%%%%%%%%%%%%%%%%%%%%%%%%%%%%%%%%%%%%%%%%%%   
%%%%%%%%%%%%%%%%%%%%%%%%%%%%%%%%%%%%%%%%%%%%%%%%%%%%%%%%%%%%%%

%%%%%%%%%%%%%%%%%%%%%%%%%%%%%%%%%%%%%%%%%%%%%%%%%%%%%%%%%%%%%%%%%%%%%%%%%   
%%%%%%%%%%%%%%%%%%%%%%%%%%%%%%%%%%%%%%%%%%%%%%%%%%%%%%%%%%%%%%

The \CCC Demonstration R\&D Plan will open up significant new scientific and technical opportunities based on development of high-gradient and high-efficiency accelerator technology. Accelerator technology that forms the basis of \CCC is presently being pursued and developed for multiple small-scale accelerator projects. The \CCC Demonstration R\&D Plan will push this technology to operate both at the GeV level and mature the technology to be reliable and provide high-brightness electron beams. The rf photo-injector for the demonstration will be a new design based on the recent advances in cryogenic copper technology with significantly higher peak field at the emission surface. These GeV-class high-brightness beams produced as part of the \CCC demonstration plan could be utilized for a variety of follow-on experiments targeting HEP applications, for beam physics studies, advanced accelerator concepts (plasma-wakefield staging, plasma lenses), positron target development, exploring novel concepts in generation of polarized beams, and developing advanced techniques for machine control with AI/ML. Already with GeV-class high-brightness beams many applications outside of HEP would be directly impacted by \CCC technology. For example, FEL, gamma-ray or x-ray inverse Compton sources could be developed. The high-brightness high-charge electron source could be utilized for ultra-fast electron diffractometry/microscopy or incorporated as a new electron source for existing x-ray FELs to push pulse energy, x-ray energy or repetition rate.

\begin{table}
\begin{center}
\begin{tabular}{c c}
     
\hline
Goal arrival time stability (fs)  &20 \\
Goal relative energy stability & $5 \cdot 10^{-4}$\\ 
Goal peak current stability  (\%)& 0.5 \\
\hline
Injector tolerances & \\
\hline
Photo cathode laser pointing stability ($\mu$m) & $\leq10$ \\
Photo cathode laser energy stability (\%) & 0.84\\
Photo cathode laser jitter (fs) & 55\\
%Charge stability (pC) & 1.68\\
S-band RF phase stability (deg) & 0.02\\
S-band RF amplitude stability (\%)& 0.04\\ 
%X-band RF phase stability (deg) & 0.033 \\
%X-band RF amplitude stability (\%) &0.2 \\
\hline
Main Linac tolerances & \\
\hline
C-band RF phase stability (deg)  & 0.3\\
C-band RF amplitude stability (\%) & 0.1\\
Quadrupole alignment ($\mu$m) & 10\\
Structure alignment ($\mu$m) & 10\\
\hline 
\end{tabular}
\end{center}
\caption{Target tolerances for electron injector and cryomodules for the \CCC demonstration.}
%For Reference: Typical tolerances for SwissFEL~\cite{PhysRevAccelBeams.19.100702,SwissFEL}}
 \label{tab:facility}
\end{table}

\section{Objectives \& Timeline}
The objective of the \CCC Demonstration R\&D Plan  is aimed at mitigating \CCC technical, schedule, and cost risks by achieving the following Technical Milestones and the Critical Parameters in Table \ref{tab:tolerance}:
\begin{itemize}
    \item High-gradient ($\geq$155~MeV/m) test with gradient margins of a damped and detuned \CCC accelerating structures at S-band and C-band
    \item Testing of the cryomodule at gradients required for \CCCfive\ (goal is to reach $\geq$155~MeV/m for margin)
    \item Demonstration of a fully engineered cryomodule with full beam loading in at least one cryomodule
    \item Demonstration of full \CCC \LN and vapor flow in 3 cryomodules as well as verification of vibrations and alignment stability.
    \item Experimental verification of the tolerances and specifications for production of the main linac cryomodules, including assembly procedures.
    \item Demonstration of a viable rotation and cooling system for a positron production target 
\end{itemize}

\begin{table}
\begin{center}
\begin{tabular}{c c}
     
\hline
\hline
Accelerating Structure & \\
\hline
Gradient (MeV/m)  & 70/120 \\
Flat Top (ns)  & 700/250 \\
Shunt Impedance (M$\Omega$/m) & 300\\ 
Breakdown Rate (/pulse/m) & 3$\times10^{-7}$ \\
Beam Loading (\%) & 45 \\

\hline
Alignment $\&$ Vibrations \\
\hline
Main Linac Components ($\mu$m)  & 10\\
Beam Delivery System Components ($\mu$m) & 10\\
Vertical Stabilization Linac Quad (nm $>$ 1 Hz) & 1.5 \\
\hline 
\end{tabular}
\end{center}
\caption{Critical Parameters}
 \label{tab:tolerance}
\end{table}

\begin{figure}[!ht]
    \centering
    \includegraphics[width=1.05\textwidth, angle =90 ]{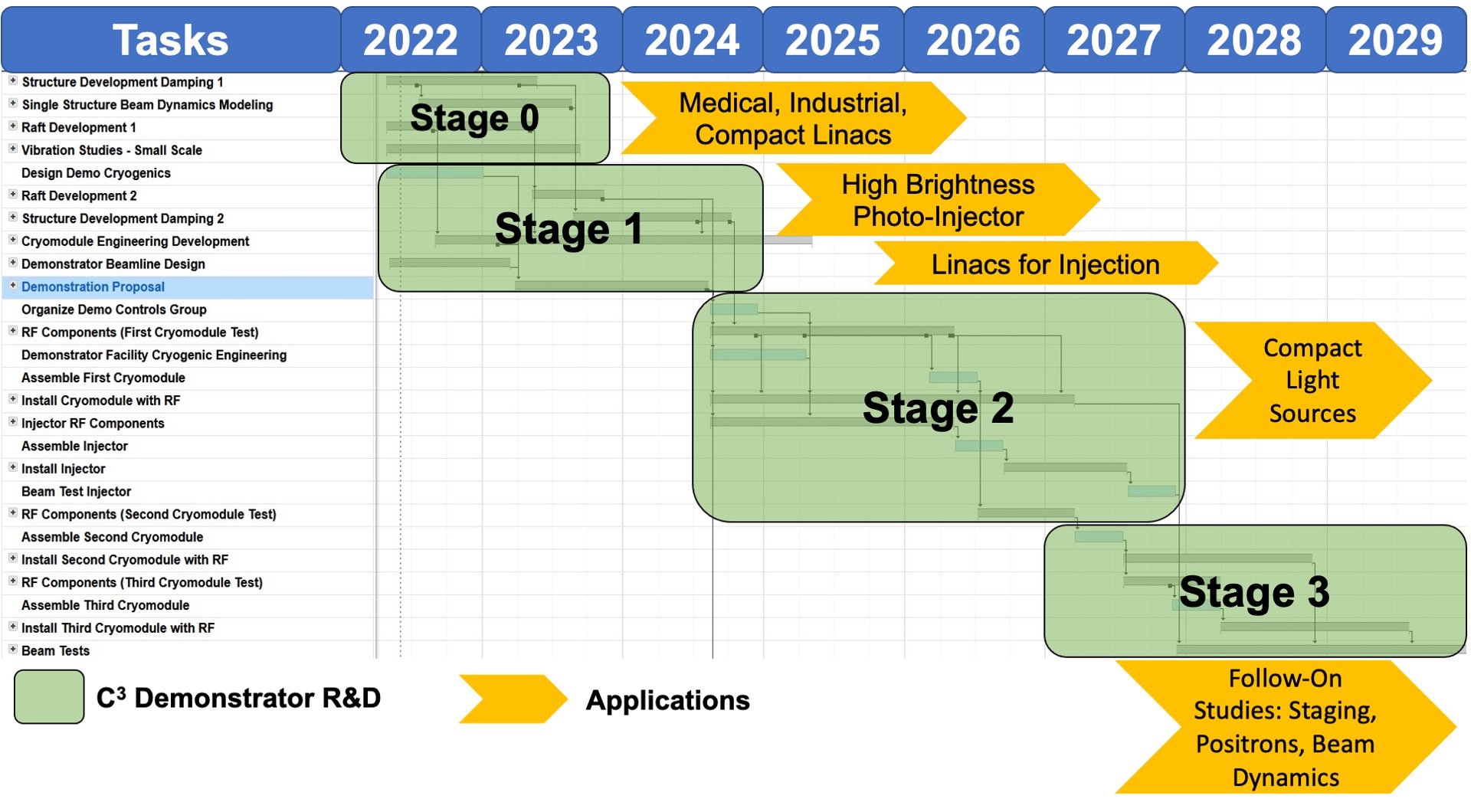}
    \caption{\CCC Demonstration R\&D Plan timeline. The timeline is undertaken in four stages. Stage 0 consists of ongoing fundamental R\&D for structure design, manufacture and test. Including operation at high gradient, damping and detuning designs, vibrations and beam test. Stage 1 is the preparatory phase for preliminary engineering studies and submission of a full proposal. Stage 2 delivers the first cryomodule, the injector and and beam tests at the full \CCC accelerating gradient. Stage 3 completes two additional cryomodules and completes beam tests. Many applications will be enable by this technology development plan and the rough timeline for those applications is also highlighted. }
    \label{fig:Timeline}
\end{figure}

The timeline for progressing with \CCC technology development will be governed by practical limitations on both the technical progress and resource availability. Here we present the technically limited timeline for the Demonstrator R\&D Plan. The detailed breakdown is shown in Figure \ref{fig:Timeline}. 

\textbf{Stage 0} At present, fundamental rf accelerator R\&D that is relevant to \CCC is ongoing with the investigation of high-gradient accelerating structures, rf design, rf components for power distribution and compression, frequency and temperature scaling, materials, manufacturing techniques, wakefield suppression, novel rf source concepts, and experimental prototypes at the meter scale. This fundamental R\&D is directly impacting the development of small-scale and compact accelerators for industrial, security and medical applications ranging from 10-100 MeV. These advances are also impacting larger scale facilities, for example with the adoption of new concepts for pulse compression and transverse deflectors. This technology is also being adopted for both the rf gun and linac for compact x-ray and FEL sources.

Through the Snowmass process we are actively exploring how these innovations in rf accelerator technology impact the design of \CCC facilities that scale from Higgs factory to a few TeV. This has allowed us to identify the key parameters and milestones for advancing the \CCC concept.

\textbf{Stage 1} for the \CCC Demonstration R\&D Plan will serve to advance the development of \CCC technology to a level of engineered maturity to form the basis for a demonstrator proposal. In particular, we will focus on the engineering of the cryomodules that would be tested as part of the demonstrator. This will include a damped-detuned accelerating structure, prototype support rafts with alignment and vibration suppression, and full consideration of the cryogenics design of the cryomodule. This will also include the design of the injector and \CCC demonstrator beamline design with start-to-end simulations including space-charge and wakefield effects. In addition to providing the technical information needed to produce a full demonstrator proposal, Stage 1 activities will directly impact applications requiring compact accelerators such as the activities described in Sections \ref{sec:UCLA} and \ref{sec:LANL} targeting compact x-ray and gamma-ray sources. The injector design could also be utilized for applications targeting very high charge electron bunch sources (for examples as an upgrade to the EIC injector\cite{nanni2021}) or the high-brightness rf photo-injector (\ref{sec:HBgun}) design could be utilized for a an upgrade to LCLS-I to produce very high energy soft x-ray pulses \cite{XCC}, as discussed in \ref{sec:XCC}. In particular, the distributed-coupling rf linac technology would also impact the realization of medium energy electron beam injectors as it provides for a resilient design against failures in the rf subsystems, i.e., klystrons. A demonstration S-band prototype accelerator structure for electron injectors is currently under construction at SLAC to demonstrate the attainable gradient, optimized beam aperture and operational reliability \cite{nanni2021}.

\textbf{Stage 1} will terminate with submission of a full proposal for the \CCC Demonstration R\&D Plan. We aim for a decision to proceed at the end of 2024. 

\textbf{Stage 2} of the \CCC Demonstration R\&D Plan will focus on the production and testing of the first cryomodule. The first engineered design for a full cryomodule will be tested initially with high power rf. The system will be operated at cryogenic temperatures. The gradients that are achieved would accelerate a beam to by 1~GeV in 9~m. Captured dark current would allow for confirmation of this energy gain and gradient. In parallel we will build and commission the electron source for the \CCC demonstrator. This will consist of a high-brightness rf photo-injector and two meters of S-band linac that can support high charge. This linac would be adopted in the electron and positron sources for \CCC up until the damping ring. After commissioning the injector, the photo-emitted beam would be compressed and transported into the first cryomodule and the first beam tests would commence. This would be a major milestone for \CCC and provide sufficient experimental data to compile a CDR, but also deliver a high-gradient accelerator technology that could be mass produced and used for compact FELs, XFELs, compact gamma ray sources, or a beam driver for PWFA. Stage 2 of the \CCC Demonstration R\&D Plan is anticipated to take 3 years (2025-2027) and concludes with the transport of photo-electrons through the first cryomodule.

\textbf{Stage 3} of the \CCC Demonstration R\&D Plan includes updates to the engineering design of the cryomodules, production of the second and third cryomodule and their installation. The \LN boiler would also be installed to allow for full thermal and vibration tests. After the installation of each cryomodule, beam tests would be performed with increasing beam current to test full beam loading. Lower charge and lower emittance beams would be used to investigate emittance growth. Stage 3 would be completed with the operation of the final cryomodule with beam at the \CCCfive\ gradient. Sufficient information would be produced mitigating \CCC technical, schedule, and cost risks required to proceed with a TDR. After Stage 3, the \CCC demonstrator could be utilized for a wide variety of R\&D related to technological improvements for \CCC, related HEP R\&D or other applications such as a compact FEL as discussed in Sections \ref{sec:parallelR&D} and \ref{sec:HEPRD}. 

Completion of the \CCC Demonstration R\&D Plan will allow us to achieve:

\begin{itemize}
    \item Demonstration of feasibility and first optimization of the \CCC linac technology
    \item Design and optimization of the linear collider accelerator complex based on \CCC technology
    \item Confirm the suitability of the \CCC beam parameters for the physics reach and detector performance 
    \item Preparation for a Conceptual Design Report (CDR)
    \item Preparation for follow-on technology development and industrialization 
\end{itemize}

%\section{Timeline} 
%\textcolor{red}{Ideal phasing of R$\&$D from single structure prototype to full demonstrator / and first steps to string test. How detailed should this be?}

\section{\CCC Demonstration Research and Development Plan Topics}
\subsection{Direct R\&D}
%\subsection{Direct Demonstrator Facility ($\geq$50 m, 120 Hz, 133$\times$1 nC)}
%\name{Names of key people to contribute, please add yourself if interested or comment/edit}

\subsubsection{Accelerating Structures for Main Linac and Source Linac}
\label{sec:accel_structures_main_source_linac}
The implementation of new approaches to  designing, building and operating a normal conducting accelerator structure has been key to the development of the technology for \CCC. The discovery process began with an \textit{ab~initio} study of cavity shapes to maximize the on-axis accelerating field while minimizing probability of breakdown. After the first optimization, some suitable cavity shapes were proposed, however the optimized shapes had very small beam irises that precluded travelling-wave cavity coupling for the  fundamental accelerating mode. As an alternative to on-axis coupling a distributed rf coupling scheme was proposed, implemented by parallel manifold rf waveguides providing side-coupling into each cavity with the proper rf phase and fraction of the inlet rf power. It was concluded that this relatively complex structure could be machined in two halves (or 4 quarters) by low-cost numerically-controlled milling machines. In addition to allowing for fabrication of the complex cavity shapes, this novel milling process results in ultra-high-vacuum (UHV) quality surfaces that need no further finishing apart from a standard copper surface etch. The final element of the proposed approach is operating the copper accelerator at a reduced temperature to increase the electrical conductivity of the material and improve the material strength and reduce probability of breakdown. The benefit of increasing electrical conductivity  is in the reduction of the required rf power, as the rf power sources are a costly and complex part of the linac infrastructure. Increased conductivity also reduces thermal stresses in the material that  result in cyclic fatigue of the material exposed to rf pulses, crystal growth, motion of dislocations  and eventually in electrical breakdown. The onset of breakdown limits the achievable gradient for the accelerating structure and sets the practical gradient limit that the main linac can be operated at without degrading luminosity.

The \CCC structures will operate in a bath of \LN cooled to a cryogenic temperature of approximately $\sim$80~K. The increased conductivity of copper at this temperature results in a shunt impedance of the optimized accelerator cavity of 300~M$\Omega$/m, six times the effective shunt impedance of the Next Linear Collider (NLC) X-band accelerating structure.

\begin{figure}[!ht]
\begin{center}
 \includegraphics[trim={0 0cm 0cm 0cm}, clip, width=0.95\hsize]{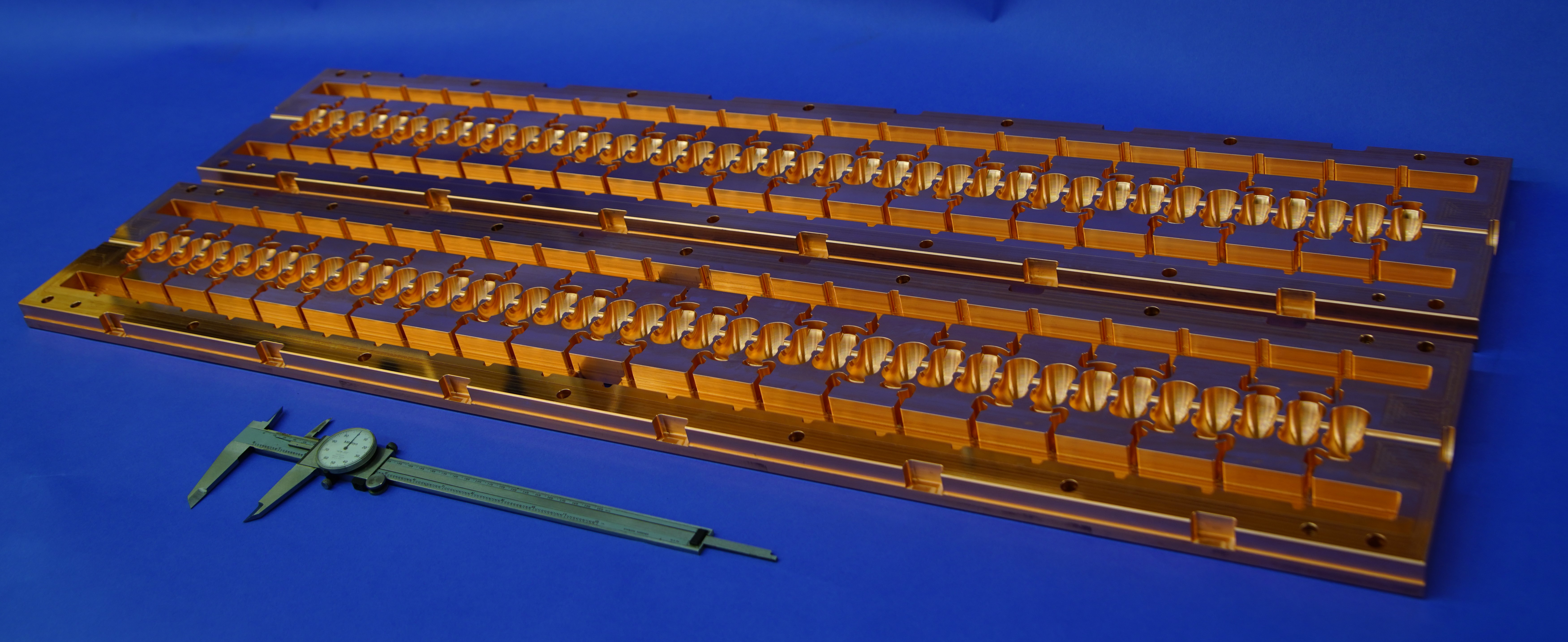} 
\end{center}
\caption{Both halves of the \CCC prototype structure prior to braze. The one meter structure consists of 40 cavities. A rf manifold that runs parallel to the structure feeds 20 cavities on each side. The structure operates at 5.712 GHz.}
\label{fig:linacstuc}
\end{figure}

\begin{figure}[!ht]
\begin{center}
 \includegraphics[trim={0 0cm 0cm 0cm}, clip, width=0.75\hsize]{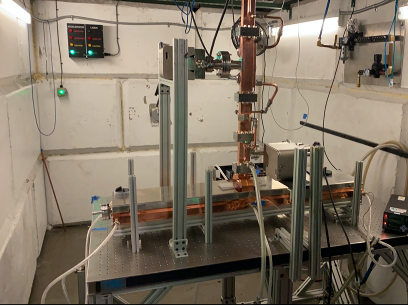} 
 \end{center}
\caption{The \CCC prototype structure in test at Radiabeam. Both room temperature and cryogenic tests will be conducted as part of the \CCC Demonstration R\&D Plan to quantify improvements to the structure performance as we incorporate damping and detuning in the design.}
\label{fig:hgtest}
\end{figure}

The \CCC accelerating structure utilizes a copper standing wave distributed coupling rf structure. The first meter-scale prototype \CCC structure is shown in Fig. \ref{fig:linacstuc}.  The aperture of the cavity is determined by considering short-range and long-range wakefield effects for the nominal bunch charge of 1~nC. The baseline phase advance between cells is $\pi$. The frequency of operation for the main linac will be 5.712~GHz (C-band) in order to provide a high shunt impedance. For the electron source, positron capture linac and booster linacs into the damping ring, the operational frequency will be  at S-band (2.856~GHz) to accommodate the longer electron and positron bunch length before compression. Prototype one meter accelerating structures have been fabricated and tested at high gradient and at cryogenic temperatures. A similar C-band accelerating structure was recently tested at Radiabeam and their room-temperature high gradient test setup is shown in Fig. \ref{fig:hgtest}.

%%%%%%%%%%%%%%%%%%%%%%%%%%%%%%%%%%%%%%%%%%%%%%%%%%%%%%%%%%%%%%%%%%%%%%%%%   
%%%%%%%%%%%%%%%%%%%%%%%%%%%%%%%%%%%%%%%%%%%%%%%%%%%%%%%%%%%%%%

%%%%%%%%%%%%%%%%%%%%%%%%%%%%%%%%%%%%%%%%%%%%%%%%%%%%%%%%%%%%%%%%%%%%%%%%%   
%%%%%%%%%%%%%%%%%%%%%%%%%%%%%%%%%%%%%%%%%%%%%%%%%%%%%%%%%%%%%%

The proposed \CCC beam format consists of trains of the electron bunches recurring at 120~Hz. Each train will have 133~bunches with the bunch charges of 1~nC separated by 5~ns for \CCCtwo, and by 3~ns for \CCCfive. This short bunch separation requires sophisticated wakefield control to prevent emittance growth. Besides, the small diameter of the iris aperture of a/$\lambda$=0.05 has the potential to result in significant short range wake effects. The emittance growth in the structure was analyzed for various beam parameters by varying the offset in phase for longitudinal wake suppression and varying the bunch length for transverse wake suppression. The goal of this simulation was to achieve a residual energy spread in the range of $\sigma_{\delta0}$ = 2-5x10$^{-3}$ and a maximum of 1$\%$ correlated energy spread for Balakin-Novokhatski-Smirnov (BNS) damping, which was analyzed with transport simulations in the main linac\cite{bane2018advanced}.

Ongoing R\&D for the structure development is focused on the design of the structure's damping and detuning to mitigate the effects of long range wakefields. The kick factor must be suppressed below 1~V/pC/mm/m to provide  stable transport of the electron and positron beams. Detuning of higher order modes is achieved by modifying the geometry of each cavity while maintaining constant frequency of the fundamental mode. The frequencies of higher order modes vary as a result of geometrical changes. Simulations have shown that $4\sigma$ Gaussian detuning with a $\Delta f/f_c$ of 6$\%$ is sufficient to suppress the longitudinal wake for the first subsequent bunch for the first dipole band~\cite{bane2018advanced}. Detuned cavity designs exist already and they can cover the necessary range in frequency ($\Delta f/f_c=6\%$). For longer range damping we are designing a structure with the use of longitudinal damping slots in quadrature that meets the required reduction in quality factor for higher order modes.

The currently existing prototype accelerating structures are sufficiently advanced to determine the rf power requirements, thermal loading, geometrical constraints and suitability of manufacturing techniques. Still, significant optimization of the accelerating structure within these limits is possible and will be the focus of early stage R\&D in this demonstration plan. Remaining within these limits will allow improvements to the accelerator structure without requiring a redesign of the cryomodule internals. These improvements to the accelerating structure are in large part enabled by the rf design of the distributed coupling topology which allows for fabrication of the structure in halves, thereby allowing for simplified incorporation of rf distribution and damping.

\subsubsection*{Accelerator Structure R$\&$D Topics:}

\begin{itemize}
\item \textbf{Manufacturing} - Minimizing part count, minimizing raw material, simplifying assembly and automating quality control and tuning. 
\item \textbf{Bonding Techniques} - Optimization of brazing with laser-cut shims, and exploration of alternative low temperature techniques with electron beam welding or diffusion bonding to increase throughput.
\item \textbf{RF Design} - Optimization of rf phase advance to $3\pi/8$ phase advance per cell, reduced cell length and proper power and phase rf manifold design for distributed coupling.
\item \textbf{Damping} - Advanced modeling and engineering are required to finalize the design of wakefield damping features in the cavity. Material loss and performance in UHV and high field environment must also be confirmed. Leading material candidates are: NiCr, FeCrAl with a special technology of coat, SiGraSiC-group composits, and  doped SiC. 
\item \textbf{Thermomechanical Analysis} - Vibration, alignment, tolerancing, and thermal analysis of accelerating structure.
\item \textbf{Enhancement of heat transfer to \LN} - Optimization of the accelerator structure surface profile to minimize N$_2$ gas bubble trapping under the accelerator structure; Exploring surface treatment for example, special texturing and coatings that enhance nucleate boiling heat transfer; surface passivation techniques that prevent surface oxidation (oxidation is known to reduce nucleate boiling heat transfer over time and so should be prevented). 
\end{itemize}

\subsubsection{Cryomodule Cryogenics}

 The accelerating structure operates immersed in a liquid bath of \LN at $\sim$80~K. The thermal load of the accelerator is cooled by the boiling of liquid nitrogen. The cryomodule transports both the liquid and gaseous nitrogen in the same vacuum jacketed enclosure that houses the accelerator. The accelerator is in cryomodules that house 4 rafts; each raft supports 2 accelerator structures, a permanent magnet quadrupole, and beam position monitors (BPM). A rendering of the \CCC cryomodule is shown in Fig.~\ref{fig:cryomod}. The improvement in Cu conductivity at $\sim$80~K improves the accelerator efficiency by a factor of 2.5-2.7X to more than recover the capital and operating expenses of the refrigeration plants. For the \CCC collider it is envisioned that the main linac sectors consist of 10 cryomodules, and super-sectors consist of 10 sectors\footnote{The actual main linacs may contain partial super-sectors.}. A super-sector is supported by a single entry and exit point for liquid and gaseous nitrogen. It is anticipated that there will be either 2 or 3 super-sectors per main linac for \CCC.

The cryomodule is a vacuum insulated tube with an inner radius of 30 cm. The accelerator components are under slowly flowing \LN at a pressure of $\sim$1.1~bar, with the \LN $\sim$25~cm above the cryostat low point. The \LN is introduced at the super-sector boundaries or the ends, and flows in both directions (one if at an end) for at most 500~m. Gaseous nitrogen counter-flows, is removed at the boundaries or ends, is re-liquified, and then re-injected. The \LN for one direction of flow enters at 7.2~l/s, and flows at a velocity (at the entry) of 0.06~m/s (0.2~km/hr). The counterflow gas flow has an area equivalent to a 23~cm radius pipe, and has a pressure drop over the full length of $\leq$10~mbar. 

The \LN flow is driven by gravity. For the \CCC collider, the spans between super-sectors are laser straight, the mid-span is normal to the earth radius there, and thus the \LN is deeper at the center of a super-sector by $\sim$7~cm. The beam will be bent in the vertical plane at the super-sector boundary by $\sim$80 micro-radians to go into the next super-sector. For a 100~GeV beam, this requires a dipole with a kick of 0.05~T-m. For \CCCtwo\ and \CCCfive, the power dissipated in one accelerator section is 2500~W, or 0.4~W/cm$^2$. This thermal load is constant because of the reduced flat top of the rf pulse. With this thermal load, the accelerator will operate in the nucleate boiling regime, and the expected temperature rise is $\sim$2~K. The temperature rise in the copper block in a 1D approximation is 0.6~K.

The open channel (“river”) of \LN flow must be driven by a pressure head, which comes from a liquid depth difference.  The various features providing liquid flow area changes, like vertical waveguides and cryomodule interconnects, result in a resistance to liquid flow and could potentially create a liquid flow problem due to a large pressure head required.  This will be carefully evaluated to provide the assured margin (through the cryomodule diameter) as the design progresses.  

Very early in the conceptual design phase, it is important to consider failure modes and the potential impact on design features.  One of those is loss of insulating vacuum, which is not unusual and often results from a leaking electrical feedthrough.  The result will be cooling and thermal contraction of the vacuum vessels, so one needs allowance for that.  Thus, not only for alignment and assembly, but also for thermal contraction, we  will investigate the inclusion of vacuum bellows periodically in the vacuum vessel string.

The \CCC Demonstrator will run in two modes:
\begin{itemize}
\item Accelerator development where there will be minimal \LN flow. The \LN will be kept at nominal height above the accelerator sections, and nitrogen vapor will be vented.
\item Vapor and liquid flow testing:
\LN flow and vapor return  between 2.5 and 10~kg/sec. The actual flow rate will be determined by the power into the \LN boiler at the end of the cryomodule string opposite to the input cross, and the input \LN flow will be adjusted to maintain the \LN height in the cryomodules.
\end{itemize}
The conceptual design is shown in Figure~\ref{fig:layout} and consists of an electron source and injector that operates at 100~MeV followed by three cryomodules. The entry point for the cryogenic liquid is between the injector and the three cryomodules.  After the three cryomodules a liquid nitrogen boiler is used to simulate the thermal load of a full super-sector with resistive heating. The boiler will consume power only during vapor and liquid flow testing.

\begin{figure}[!ht]
\begin{center}
 \includegraphics[trim={0 0cm 0cm 0cm}, clip, width=0.95\hsize]{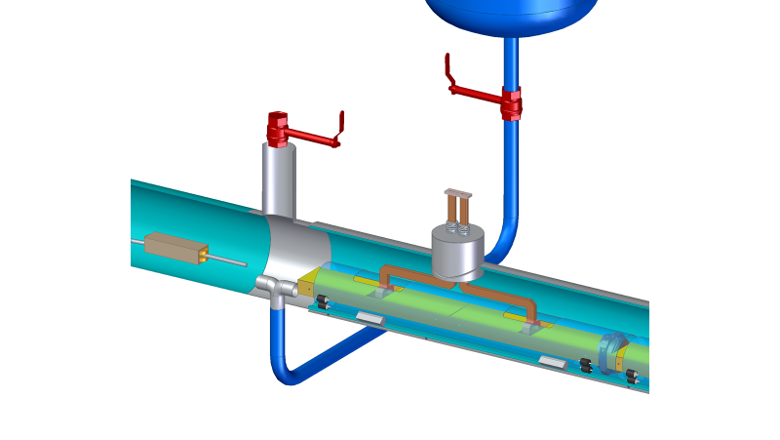} 
 \end{center}
\caption{Rendering of the cross for inserting liquid nitrogen and removing nitrogen gas from the cryomodule. Liquid nitrogen comes in from bottom and is directed laterally in two counter propagating directions. The inlet control valve regulates on liquid level and is capable of supplying the full liquid flow rate envisioned for the \CCC main linac. The outlet valve regulates on the pressure of vapor in the cryomodules.}
\label{fig:cross}
\end{figure}

\LN enters and vapor exits through a cross that will be a full prototype of a super-sector cross, see Fig. \ref{fig:cross}. \LN enters the bottom of the cross through a metering valve from a $\sim$50,000~liter \LN storage dewar kept at $\sim$80~K. The valve is controlled by the liquid level in the cryomodules, and feed forward control based on power dissipated in the liquid can be tested. Vapor exits from the top of the cross through a metering valve that is controlled to maintain $\sim$1.1~bar in the cryomodules.

%%%%%%%%%%%%%%%%%%%%%%%%%%%%%%%%%%%%%%%%%%%%%%%%%%%%%%%%%%%%%%%%%%%%%%%%%   
%%%%%%%%%%%%%%%%%%%%%%%%%%%%%%%%%%%%%%%%%%%%%%%%%%%%%%%%%%%%%%

\LN fills the boiler at the downstream end of the cryomodules to the level in the cryomodules. During the first mode of operation, the heaters are off. For the second mode of operation, the heaters can be powered up to $\sim$2~MW, which will produce $\sim$10~kg/sec of vapor. This will provide adequate margin for testing the system beyond expected flow rates.

We will pursue optical studies of nitrogen two-phase flow in the cryomodule to probe probe two-phase instabilities in the vessel. $N_2$ gas phase velocities corresponding to 10 kg/s mass flow can be high enough to initiate a surface wave instability on the \LN underneath the vapor. An optical setup such as a transparent windows on the demonstrator vacuum and \LN vessels will help probe two-phase instabilities vessel. Alternately, cryogenic cameras can be housed inside the \LN vessel to capture images of \LN surface and determine gas velocities that lead to the instability. During a cold rf test, this optical setup will also provide information on nucleate boiling, for instance whether gas bubbles actually trap under the accelerator structure.

The S-band accelerator for the the electron source of the demonstrator will also be in a cryomodule, and will be linked to the super-sector cross by a cryomodule containing the normal conducting bunch compressor. This will allow for the electron source linac topology to also be tested on the timeline of the \CCC Demonstration R\&D Plan.

\subsubsection{Cryomodule Internals Design}

The Cryomodule functionality is:
\begin{itemize}
    \item Provision of stable support for the accelerator sections and quadrupoles.
    \item Provision of the ``cool" environment of $\sim$80~K, including heat removal from the accelerators.
    \item Provision of alignment capability within a cryomodule, and to adjacent cryomodules.
    \item Provide feedthroughs for waveguides and other cables.
    \item Enabling of the above safely and reliably.
\end{itemize}

\begin{figure}[!ht]
%\begin{center}
 \includegraphics[trim={0 0cm 0cm 0cm}, clip, width=0.95\hsize]{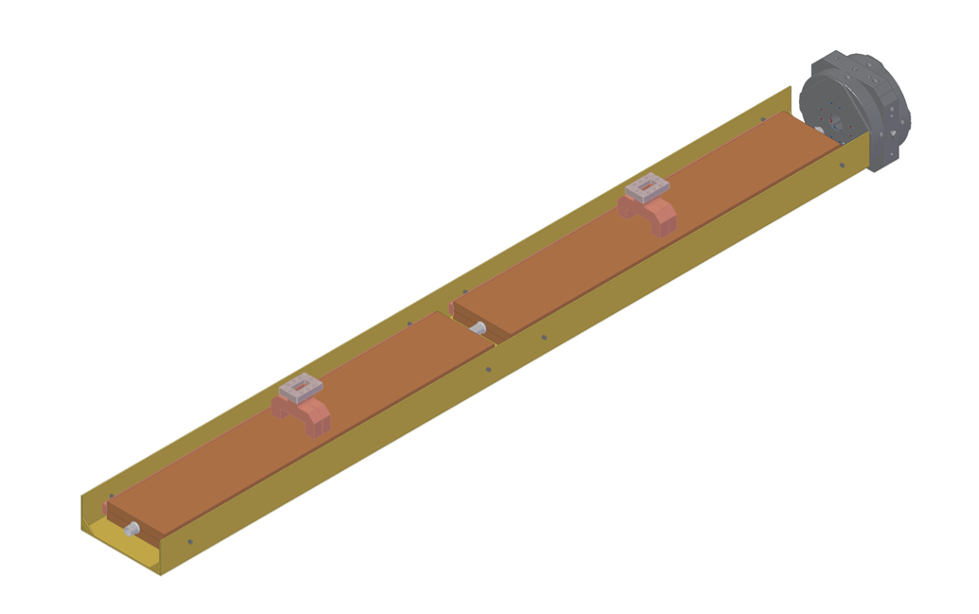}
% \end{center}
\caption{Rendering of the 2~m \CCC support raft with 2 one meter structures and one permanent magnetic quadrupole.}
\label{fig:raft}
\end{figure}

The basic mechanical building block is the ``raft", see Fig. \ref{fig:raft}. A raft consists of two accelerator sections and a permanent magnet quadrupole. The quadrupole has an integrated cavity beam position monitor (BPM). The quadrupole design shown in Fig. \ref{fig:raft} is based on a prototype developed by Electron Energy Corporation (EEC) with an innovative field adjustment mechanism using dynamically tunable magnets. EEC’s design is engineered for immersion in \LN and enables a compact and uniquely flexible permanent magnet quadrupole that achieves the field requirements, superior to current designs, for \CCC. There are 4 rafts per cryomodule which is $\sim$9~m long. There are 10 cryomodules per sector, which is $\sim$100~m long. Finally there are 10 sectors per super-sector, which is $\sim$1~km long. The \CCC Demonstration R\&D Plan calls for three cryomodules to be tested in series. The cryomodule will fit in a standard ISO shipping container (40 feet long).

%%%%%%%%%%%%%%%%%%%%%%%%%%%%%%%%%%%%%%%%%%%%%%%%%%%%%%%%%%%%%%%%%%%%%%%%%   
%%%%%%%%%%%%%%%%%%%%%%%%%%%%%%%%%%%%%%%%%%%%%%%%%%%%%%%%%%%%%%

\begin{figure}[!ht]
\begin{center}
 \includegraphics[trim={0 0cm 0cm 0cm}, clip, width=0.95\hsize]{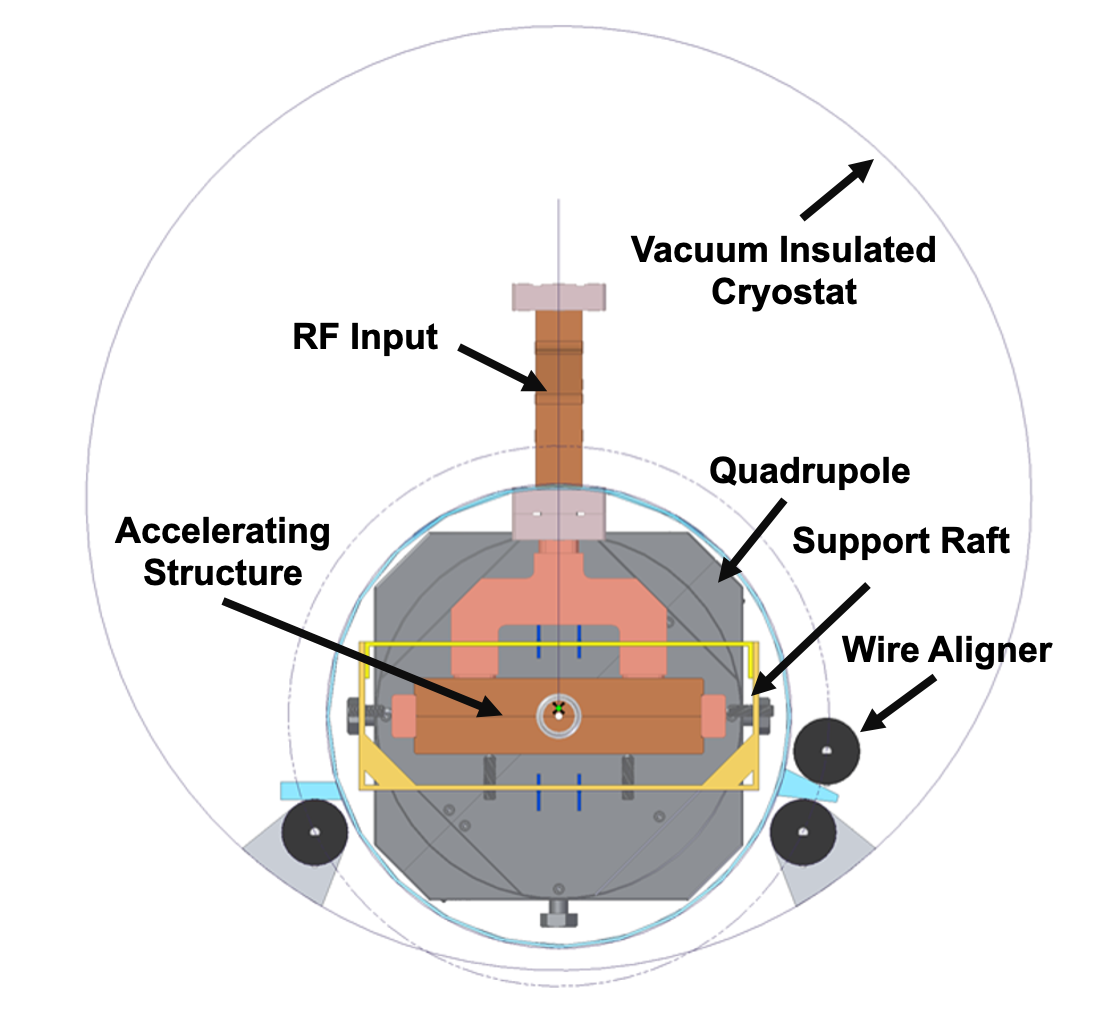}
\end{center}
\caption{Rendering of the cryomodule cross section. In this view the the accelerating structure, the waveguide feed, support raft, quadrupole, mechanical and piezoelectric movers, and stretched wire aligners are visible.}
\label{fig:cryoint}
\end{figure}

The raft components are held by a channel which is a very open truss structure. The two accelerators and the quadrupole are pre-aligned on the raft. These components have 5 degrees of freedom (d.o.f.), with $Z$ constrained by a ball in a groove at each midpoint. The raft is suspended from piezoelectric transducer stacks with a range of 200 microns and a bandwidth of $\sim$100~Hz. These actuators are for beam based alignment and dynamic feedback. The transducers are an intermediate stage suspended from high resolution cryogenic electric motor jacks, with a range of $\sim$2~mm. These jacks are primarily for inter-raft alignment, including the cryomodule transitions. It is expected that they would rarely be used after initial alignment unless there are upsets such as earthquakes. The motors are unusual in that they can be dirty in the sense of throwing off particulates, and only need a minimal bearing lifetime. A cross section of the cryomodule showing the accelerating structure, the waveguide feed, support raft, quadrupole, mechanical and piezoelectric movers, and stretched wire aligners is shown in Fig. \ref{fig:cryoint}.

%%%%%%%%%%%%%%%%%%%%%%%%%%%%%%%%%%%%%%%%%%%%%%%%%%%%%%%%%%%%%%%%%%%%%%%%%   
%%%%%%%%%%%%%%%%%%%%%%%%%%%%%%%%%%%%%%%%%%%%%%%%%%%%%%%%%%%%%%

The cryostat is a vacuum insulated pair of concentric cylinders, with a 30~cm  inner radius of the cold cryostat. The cold cryostat will (probably) be made of 304 stainless steel, which closely matches the integrated thermal contraction of Cu. The outer cryostat can be soft steel. The inner cryostat will be wrapped with Multi-Layer-Insulation (MLI). Each accelerator section is fed by one waveguide, and the two waveguides from one raft are routed together through a port. That port also accommodates cables for the alignment motors, transducers, alignment sensors, the BPM signals, and the quadrupole controls.  

A pre-conceptual plan for the assembly of the waveguide ports exists. It is hoped that all connections associated with cryomodule are welded, avoiding all bolted flanges. The weld preps would be designed for cutting (``can opener") or grinding in case access were needed.

The integrated thermal contraction from room temperature to $\sim$80~K is -0.31\% for Cu and -0.29\% for SS304. Therefore the $\Delta L$ between 1~m of Cu accelerator and 1~m of SS304 raft is $\sim$200~$\mu$m. Thus the accelerator midpoints are pinned to the stainless and the accelerator sections separate by 200~$\mu$m. This $\Delta L$ is taken by bellows on the beam pipes between accelrating structures.

The 9~m SS304 inner cryostat contracts $\sim$3~cm. Since the outer steel shell stays warm, sliding support is needed. The adjacent cryomodule inner cryostats separate by 3~cm. This $\Delta L$ is taken by bellows between the the two shells. The outer cryostats do not require bellows.

\begin {table}[h]
\begin{center}
\begin{tabular}{c c c c c c c c}
     
\hline
 Gradient & Power diss. & rf flat top & Pulse & Comments & Power/area & $\Delta T$ Cu-bulk  \\
 (MV/m) & (W) & (ns) & compr. & & (W/cm$^2$) & to \LN (K) \\
 \hline
70 & 2500 & 700 & N & \CCCtwo & 0.393 & 2.3 \\
120 & 2500 & 250 & N & \CCCfive & 0.393 & 2.3 \\
155 & 3900 & 250 & N & \CCCfive\ in 7 km & 0.614 & 2.5 \\
120 & 1650 & 250 & Y & \CCCfive\ & 0.259 & 2.1 \\
\hline

\end{tabular}
\end {center}
\caption {rf power losses in the accelerator structures for 4 considered \CCC scenarios.}
\label{tab:RF Losses}
\end {table}

The rf thermal loads are shown in Table~\ref{tab:RF Losses}. For \CCCtwo\ and \CCCfive, the loads are 2500 W per accelerator section. If \CCCfive\ is forced into 7 km, then the load will go up to 3900 W per section. If rf pulse compression is added, the thermal load can be reduced to 1650 W per section. A goal of the Demonstrator is to show that this range of loads can be handled without causing vibration problems at the flow rates required for full super-sector flows, i.e. going up to 10 kg/sec of Nitrogen.  These power dissipation values correspond to loads between 0.26 and 0.61 W/cm$^2$, which are in the nucleate boiling regime. The expected $\Delta T$ is below 2.5~K.

\subsubsection{Cryomodule Assembly}

It seems likely that the cryostat consisting of cold and warm tubes plus the wave guide ports and and inter-module connections can be commercially procured for the Demonstrator. The cryostat would also be fitted with the sliding supports for contraction of the cold tube. This cryostat would be leak tested warm and shipped to the assembly site in an ISO container fitted with suitable restraints. 

Rafts will be assembled and aligned on a 5 m granite table placed so the rafts can be coupled together and pulled into the cryostat along with their alignment wires. It will be possible to adjust the alignment wires by reaching through the waveguide ports. All the remote alignment systems (motor driven jacks and piezo transducers) will be tested before raft insertion. Each raft will have a locking feature to the cold cryostat for transport that is accessible from the cryostat ends. These locks will be released before final placement of the cryomodules on the beamline.

\subsubsection{Beam Dynamics}

There are some key beam dynamics studies required to fully validate the new style of distributed coupled structures. Following a program similar to NLC in the past, the impact of short and long range wakefields on the accelerated electron and positron beams need to be studied in comprehensive start-to-end tracking simulations, with associated experimental tests at the demonstration facility to qualify the expected wakefield kick factors. Such an experimental demonstration would closely follow measurements made for the NLC program at the NLCTA test facility~\cite{nlcta}. A magnetic spectrometer following the \CCC accelerating structures can be used to determine the beam energy and energy spread and bunch-to-bunch offsets. A kicker in the out-of-bend plane can provide the ability to separate the bunches to allow independent measurement along the bunch train. Optics before the chicane will be implemented, along with profile measurement devices, to infer the emittance of the beam.

Vibration and alignment tolerances are also a key consideration. Again, tolerance calculations will follow from start-to-end simulations, including the beam delivery system, which include such dynamic error sources and associated beam-based feedback systems. Experimental validation of the alignment of structures and magnets, together with direct measurement of structure vibrations and the direct impact on the measured beam properties at the demonstration facility will directly feed into these simulations. The demonstration facility beam measurement suite will also be utilized to directly measure the impact of observed breakdown events in the structures.

\subsubsection{DC Polarized Electron Gun and Injector}

The baseline electron (polarized) and positron (unpolarized) sources for \CCC are conventional linear collider designs. For the electron source, this consists of a polarized DC gun, buncher and accelerator. Extensive development has been undertaken for a DC polarized electron source that is able to meet the requirements for a linear collider (including the parameters required \CCC)\cite{poelkercebaf,rinolfi2011clic}. The baseline plan for the \CCC demonstration does not incorporate the testing of a DC electron source with the cryomodules. However, as the design of the \CCC collider progresses, the DC guns that have been designed will need to be revisited for the specific parameters of \CCC. If it is deemed necessary to test the acceleration of the beam to a few GeV, the \CCC demonstration facility would be a suitable facility for doing so. In addition to a DC gun as the electron source for \CCC, we are also exploring the possibility of a polarized rf photo-injector  and it would be the target for this demonstration plan to incorporate such a (unpolarized) photo-injector first.

\subsubsection{RF High-Brightness Photo-Injector for the Demonstrator}
\label{sec:HBgun}

The significant increase in peak surface fields, upwards to 250 MV/m, with cryogenic copper accelerating structures also opens a new frontier in the development of high brightness polarized electron sources \cite{rosenzweig2018ultra}. The achieved surface fields in cryogenic copper structures are in principal sufficient to produce an asymmetric emittance low enough that it would negate the need for the electron damping ring \cite{robles2021versatile}. Indeed this source is under development at UCLA as is discussed in Section \ref{sec:siting}. Early stage R\&D for this cryo-RF electron gun is underway, so that a source of this type could be utilized for the demonstration facility. The rf gun for the \CCC demonstrator would operate at S-band to support high charge bunches. The injector for the demonstrator linac will consist of an rf photo-injector with a removable CsTe cathode~\cite{PhysRevSTAB.18.043401}. The goal will be to extract 133 one nC bunches from the photo-injector. The photo-injector will operate at S-band (a sub-harmonic of the C-band accelerator frequency). The photo-injector will produce 3-5~MeV electron bunches. For the demonstrator a symmetric emittance  electron bunch will be sufficient. We will not implement a polarized photo-cathode in the initial operation of the rf gun at the demonstrator. The development of this electron photo-injector has direct implications for high-energy XFELs and $\gamma\gamma$ collider concepts\cite{XCC} as discussed in \ref{sec:XCC}.

\subsubsection{RF High-Brightness Polarized Photo-Injector}
\label{sec:HBpolarizedgun}

Several co-authors of this white paper (Patterson, Liepe, Maxson, Musumeci, Rosenzweig)  also participate in the Center for Bright Beams (CBB), a multi-institutional National Science Foundation Science and Technology Center focused on high brightness electron beam science. The center is composed of three research thrusts: one focused on advancing the brightness of photocathode sources, another on improving the performance of superconducting RF cavities, and a third on developing methods for advanced beam control and fundamental understanding of high brightness beam dynamics.  Several aspects of the Center’s research, as well as other related sponsored projects of Center PIs, are directly relevant to \CCC R\&D, as described below; significant infrastructure and expertise in these areas are already in place.   

All proposed future $\ee$ colliders including \CCC require an electron source with a high degree of spin polarization.  The current state-of-the-art polarized electron source remains GaAs activated to negative electron affinity with a sub-monolayer of Cs and O$_2$ or NF$_3$. Several critical advancements have improved performance beyond that of bulk GaAs: superlattice structures have been shown to both dramatically enhance electron spin polarization\cite{maruyama2004systematic}, and the inclusion of a Fabry-Perot-style optical resonator via additional layered structures significantly enhances quantum efficiency by increasing absorption\cite{liu2016record}.  However, perhaps the most challenging aspect using GaAs photocathodes still remains: the Cs-containing activating layer is extremely sensitive to oxidation, preventing use in all but extreme high vacuum (P\textless1e-10 torr). 

CBB research in photocathodes devotes significant effort to the growth and characterization of high quantum efficiency semiconductor materials. Among these, both Cs$_3$Sb and Cs2Te have been shown to be able to activate GaAs to negative electron affinity. When grown thin enough, this activating layer preserves the high electron spin polarization of superlattice GaAs\cite{bae2020improved}. Critically, while both Cs$_3$Sb and Cs$_2$Te must also be operated in pure vacuum, they are far more robust than the previous monolayer Cs-based coatings; in low-voltage test chambers, these spin-preserving activating layers have been shown to have significantly longer lifetimes\cite{cultrera2020long}. 

The cryocooled copper technology that underpins \CCC enables the possibility of a very high field photocathode rf gun (photocathode field in excess of 240 MV/m), with a brightness that potentially eliminates the need for an electron damping ring. As the lifetime of polarized GaAs photocathodes is largely determined by the integrity of the activating layer, it will be critical to test these longer-lifetime activating materials in high extraction field for lifetime, dark current, and intrinsic emittance. CBB researchers are actively collaborating to test photocathodes grown at Cornell in the high field photocathode rf guns at UCLA. CBB also actively investigates other novel photocathode protection coatings, such as 2D materials, which may further improve lifetime in extreme accelerator environments. 

CBB researchers also actively investigate the performance of cryogenic rf cavities. The primary cryogenic testing infrastructure for CBB is housed at Cornell University. Beyond CBB,  the Cornell SRF group has decades of experience with in-house design, testing, and construction of superconducting RF cavities and full cryomodule cooled to 2 K. The Cornell SRF group has also pioneered several novel material approaches to cryogenic rf cavities, including demonstrating superconducting Nb$_3$Sb cavity coatings with very high quality factor which can operate at significantly higher temperatures than Nb-only SRF cavities. Dark current is a critical limiting factor for accelerating structures operating at the extreme gradient limit. Cornell is well-suited for the investigation of surface preparation techniques that mitigate dark current at cryogenic temperatures, as well as surface coatings that passivate field emission or increase the quality factor of the structure, thereby limiting the energy dissipated in the cavity thus potentially the breakdown rate. 
There are also CBB-sponsored activities at UCLA, embracing mainly ultra-high field and cryogenic cavity issues, as is discussed in Section \ref{sec:HBgun} and \ref{sec:siting}.

\subsection{High Power Klystron Upgrades}
The high power rf system for the initial demonstrator will consist of commercially available 50~MW C-band klystrons and modulators; 18 klystrons are required. To provide a path for full scale implementation, it is necessary to adopt the latest innovations for rf source design leading to increased rf source efficiency and power at equivalent or reduced construction costs. Indeed, for a full-scale collider, the fabrication and installation costs of the high power rf system can be comparable to that of the tunnel and accelerating structure itself.
In the near term, efficiency improvements can be had by designing an optimized rf circuit. Techniques for maximizing klystron efficiency, such as the Core Oscillation Method (COM), were explored in depth through the High Efficiency International Klystron Activity (HEIKA), yielding klystron efficiencies up to 80$\%$ at X-band; these same methods could be deployed at C-band.\cite{cai2021XBandKly} Fabrication and operation costs may be reduced further by replacing the conventional solenoid focusing with a periodic permanent magnet (PPM) focusing structure that leverages existing mass-produced magnets. Efficiency can be increased further by incorporating PPM focusing within a multi-beam device to reduce the effective perveance of the device. Finally, advances in commercially available additive manufacturing will be evaluated for rf sources, and may be incorporated in low-risk assemblies (for example, in “dressing” assemblies outside the vacuum envelope) when such an approach can lead to substantial fabrication cost reductions.

\subsubsection{Low Level RF and Klystron Controls}
The rf phase requirements of 0.1\%  and 0.3\degree C-band phase (150 femtoseconds) are comparable to the requirements of 4\textsuperscript{th} generation light sources. A recently developed Low Level rf (LLRF) system at SLAC provided \textless 20 femtosecond RMS drive noise and \textless 5 femtosecond RMS readback noise in a 1 MHz bandwidth, considerably better than required for \CCC. The klystron modulator interlocks are comparable to those on existing accelerators.

Neither the LLRF or rf controls require any new technology in order to meet the \CCC requirements, and any development would be directed at reducing cost. As the underlying technology is rapidly evolving, detailed designs for both the \CCC demonstration and the collider should be delayed until there is a firm project schedule in order to allow use of the most recent technology.   Its is likely that these systems will be based on highly integrated RF / FPGA / CPU parts such as the Xilinx RFSOC. 

\subsubsection{Precision Timing / Phase Distribution}
This is required in order to control the phase of the rf stations.  Again the requirements are similar to  those for 4\textsuperscript{th} generation light sources and long distance phase stabilised rf over fiber systems are available commercially. The long distance fiber systems can be combined with local reflectometer based rf phase distribution similar to that used on LCLS-II. As with LLRF, this is a rapidly evolving technology and the design should be deferred until there is a firm project schedule. Systems to phase lock rf stations and mode locked lasers to rf phase references meeting these requirements have been used in accelerator / FEL systems at various laboratories.

\subsubsection{Beam Diagnostics}
The beam diagnostics resolution requirements for \CCC are similar to those for 4\textsuperscript{th} generation light sources, but with the additional requirement of closely spaced (few nanosecond) bunches. 

Cavity BPMs are able to provide both high spatial resolution and with high bandwidth electronics can provide independent measurement of bunches in the train. Typically cavity BPM measurements are limited by mechanical stability rather than by the readout electronics. Note that the readout electronics could be identical to the LLRF electronics if high bandwidth RFSOC devices are used.  
Multi-bunch cavity BPMs operating under conditions where the bunch spacing is \textless ~cavity decay time, have been tested, but not used in production.  The algorithm is straight forward:  Measure the change in vector field amplitude caused by each additional bunch in the train.  There are tradeoffs between using cavities that are resonant with bunch rate which provide better train average position resolution, and using off-resonance cavities which provide better single bunch resolution. 

Beam profile monitor technology will be driven by details of the accelerator design. Optical Transition Radiation monitors (OTR) provide high spatial resolution images, but cannot be used with beams with high longitudinal brightness due to coherent effects. In general they will function after a damping ring but not with a beam from an rf gun. The alternative of wire scanners can be applied anywhere beam densities are not too large, but they provide only integrated, multi-bunch scans and are very slow.   In areas where the charge density will destroy any type of intercepting diagnostic, laser wire scanners are the only known option. 

Bunch arrival timing can be determined with rf pickup cavities that operate similarly to a BPM cavity. If a precision phase reference is provided, each BPM reference cavity can also serve as an arrival time monitor.  This type of system has demonstrated \textless 10~femtosecond single pulse noise, easily sufficient for this application.

\subsubsection{Accelerator Raft Alignment}

The \CCC accelerating structures are supported by rafts. The rafts are responsible for providing a mechanism for alignment in warm and cold states with mechanical actuators and with piezoelectric feedback for beam based alignment. Each cryomodule contains 4 rafts; each raft supports two 1~m long accelerator sections and a quadrupole with mechanically integrated BPM. The accelerators and quads are attached to the raft with 5 d.o.f. ball joint adjustable tie rods, and are aligned on the bench before insertion into the cryostat. The $Z$ position of the midpoint is mechanically fixed.

\begin{figure}[!ht]
\begin{center}
 \includegraphics[trim={0 0cm 0cm 0cm},angle =90, clip, width=0.5\hsize]{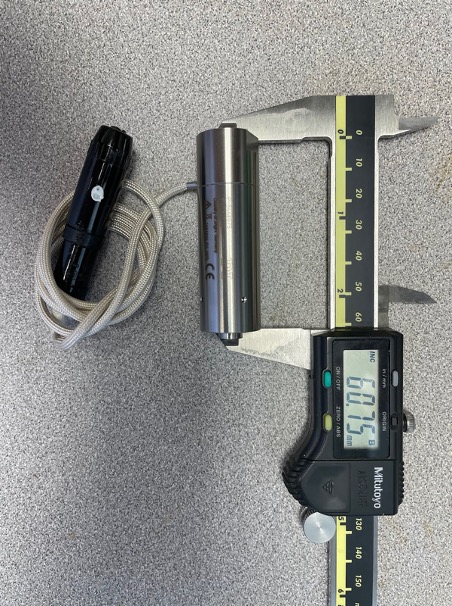} 
\end{center}
\caption{A 200 micron travel piezoelectric actuator. Actuators operate at cryogenic temperatures and high mechanical loading. The devices were developed for LCLS-II.}
\label{fig:actuator}
\end{figure}

The rafts will have coarse and fine alignment, with each raft having 5 d.o.f. ball joint adjustable tie rods. The $Z$ position of the midpoint is fixed. Each tie rod will have both a 200 micron travel piezoelectric actuator as shown in Fig. \ref{fig:actuator}, and a magnetically coupled manual screw adjustment from outside the cryostat. The fine alignment of the rafts will be beam-based and dynamic with a bandwidth of $\sim$100~Hz.

%%%%%%%%%%%%%%%%%%%%%%%%%%%%%%%%%%%%%%%%%%%%%%%%%%%%%%%%%%%%%%%%%%%%%%%%%   
%%%%%%%%%%%%%%%%%%%%%%%%%%%%%%%%%%%%%%%%%%%%%%%%%%%%%%%%%%%%%%

The coarse alignment of the rafts is done with stretched wires. Each wire is $\sim$20~m long and spans two cryomodules, with the wires starting on successive cryomodules. Each raft has two wire position sensors for each wire, one on the first accelerator and the other on the quad. The sensors measure $X$ and $Y$, with a precision of $\leq$100~microns. The rafts are slightly over-constrained with 8 measurements for 5 d.o.f.

The stretched wire method\cite{coosemans1999performance} is chosen since it must operate in air, vacuum, and under liquid nitrogen (rf off). Early-on in the demonstrator R\&D plan stretched wire prototypes must be designed and tested for operation in liquid nitrogen, and to confirm that their performance is not degraded in this environment.

\subsubsection{Vibrations}

\begin{figure}[!ht]
\begin{center}
 \includegraphics[trim={0 0cm 0cm 0cm}, clip, width=0.95\hsize]{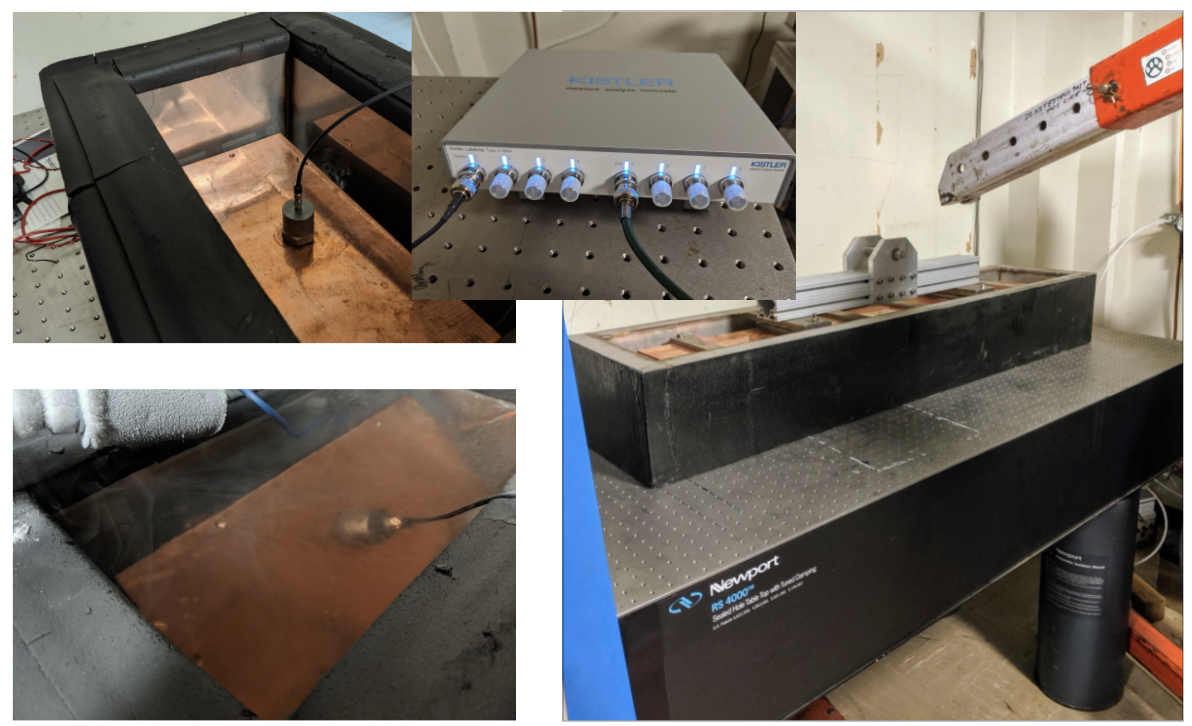}
\end{center}
\caption{Test bed for measuring vibrations from the boiling of liquid nitrogen due to applied thermal load in copper structure.}
\label{fig:vibtestbed}
\end{figure}

As noted, three stages of alignment are foreseen: initial alignment of the accelerators and quadrupole on the raft; alignment of the rafts with respect to each other by stretched wires; and finally beam based alignment. It is also required that structure vibrations be kept to acceptable levels. Vibrations can be driven by seismic disturbances\cite{decker1998status}. Prospective sites should be adequately quiet, and this has been studied by NLC, ILC, and CLIC. A new vibration source needs to be studied that arises from the nucleate boiling of liquid nitrogen and possibly the flow of nitrogen vapor. Preliminary measurements were started with a dummy accelerator fitted with 3 cryogenic accelerometers \cite{kistler} along the 3 axes, with everything under LN$_2$, and a small heater in the Cu slab. The test bed for experimenting with vibrations induced from thermal loads is shown in Fig. \ref{fig:vibtestbed}. 

%%%%%%%%%%%%%%%%%%%%%%%%%%%%%%%%%%%%%%%%%%%%%%%%%%%%%%%%%%%%%%%%%%%%%%%%%   
%%%%%%%%%%%%%%%%%%%%%%%%%%%%%%%%%%%%%%%%%%%%%%%%%%%%%%%%%%%%%%

The continuation of this study should  be done before substantially more design work is completed. Bubble formation and particularly bubble aggregation under the accelerator at high thermal loads should be studied to understand margins. This should utilize  direct optical observation of the accelerator surfaces as well as accelerometer measurements.

\subsubsection{High Power RF Distribution}

\begin{itemize}
   % \item C-Band waveguides (WG),
    %\item WG networks,
    %\item RF WG monitors,
    %\item Phase adjustments,
    %\item Beam induced monitors
    %\item RF windows and RF gaskets and other WG components including the WG vacuum valve 
    %\item Issues of klystron replacement and WG vacuum
    %\item WG cooling issues, WG temperature stability, phase length consideration in the WG network (including the RF power splitters, bends, etc.)
    %\item RF loads 

\item\textbf{Specification of waveguide network}

This section describes the high power rf distribution system at C-band (5.712~GHz). The system transmits rf pulse power from the klystrons to accelerating structure. The rf distribution system will provide for monitoring the pulse power level of the klystron output. The major specifications of the waveguide network are shown in Table \ref{tab:wgdist}.

\begin {table}[h]
\begin{center}
\begin{tabular}{c c c c}
     
\hline
Operating frequency (MHz) 5712  \\
\hline
Waveguide Scattering Coefficient (S11) less 0.07  \\
\hline
RF peak power (MW) 80   \\
\hline
RF average power (kW) 20 \\
\hline
Phase deviations of output RF power relative to an input reference phase (deg) less than 3 \\
\hline
Vacuum operating level (Torr) 10x10$^{-7}$ at full rf power and 10x10$^{-8}$ at zero rf power  \\
\hline
Gas burst rate (Rate) Less than 1 burst per hour average at full RF peak power after conditioning  \\
\hline

\end{tabular}
\end{center}
\caption{Specifications for the Waveguide Power Distribution Network}
\label{tab:wgdist}
\end{table}

An initial condition for the design of the waveguide network is based on the decision to feed each standing wave structure raft from one klystron. In this case, the first step of the rf phasing can be made at a low level rf drive signal and less accuracy is needed for the waveguide temperature stability. The precision adjustment of the accelerating rf phase is performed by a phase shifter installed in each waveguide branch. A simplified rf network layout is shown in Figure \ref{fig:Network}. The rf network layout is indicated with flanges connecting hardware components. Evaluating which of these we can replace with welds will be one of the activities for the \CCC Demonstration R\&D Plan.

%%%%%%%%%%%%%%%%%%%%%%%%%%%%%%%%%%%%%%%%%%%%%%%%%%%%%%%%%%%%%%%%%%%%%%%%%   
\begin{figure}
\begin{center}
 \includegraphics[trim={0 0cm 0cm 0cm}, clip, width=0.75\hsize]{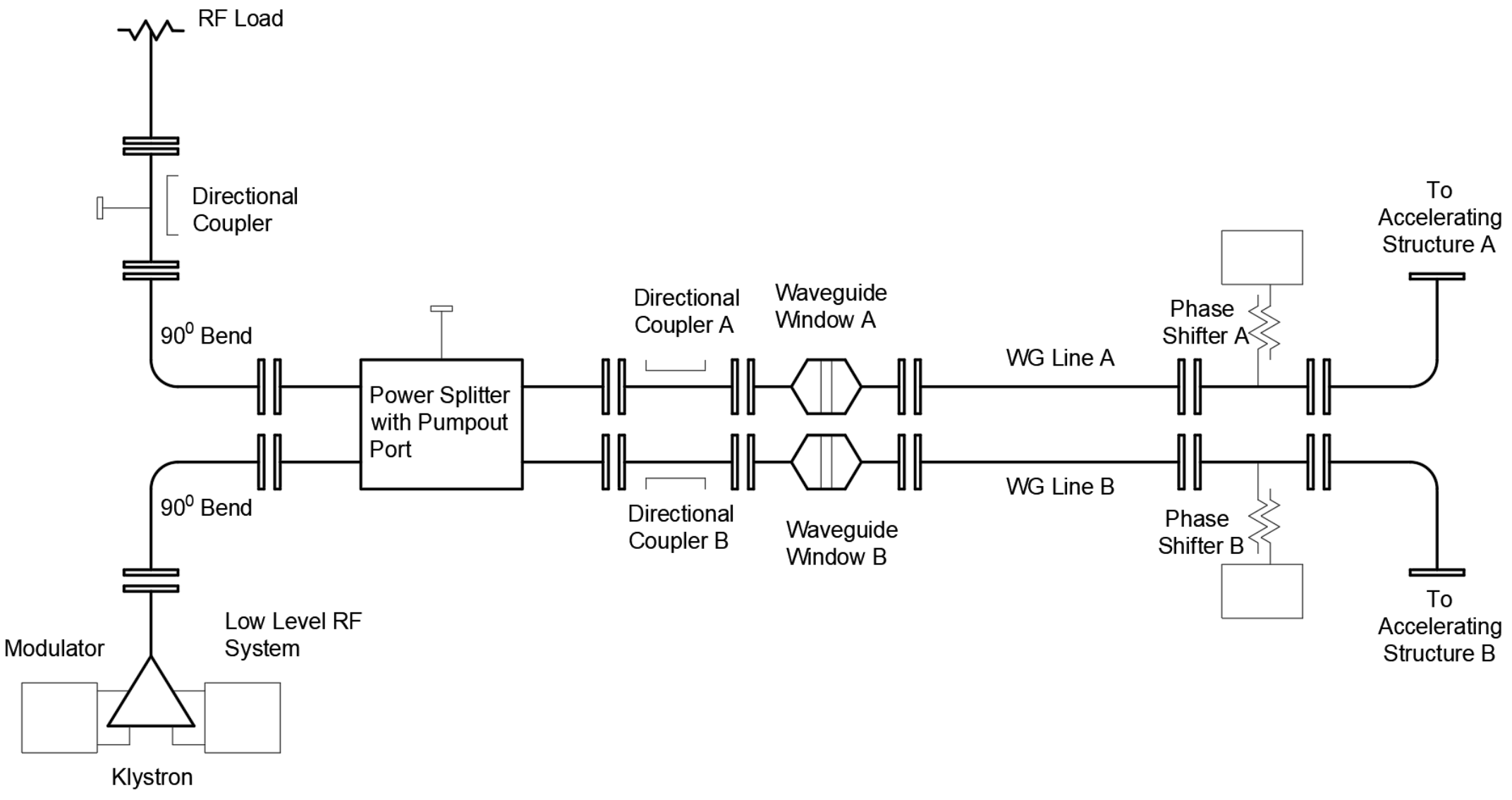}
\end{center}
\caption{Schematic of the high-power rf waveguide distribution network. The configuration shown is for one klystron feeding a 2~m raft.}
\label{fig:Network}
\end{figure}
%%%%%%%%%%%%%%%%%%%%%%%%%%%%%%%%%%%%%%%%%%%%%%%%%%%%%%%%%%%%%%

The waveguide system will be designed to operate in high vacuum. This decision dictates developing of reliable vacuum directional couplers, power splitters, vacuum pumpout orifices with a high conductance and negligible rf coupling. The primary elements affecting the waveguide layout will also include the mechanical sub-system allowing rapid replacement of the klystron. The overall \CCC rf distribution design philosophy is targeted at allowing the replacement of the klystron without significant work at the waveguide network.

Three standard waveguides can be used to transmit RF power from the klystron to the accelerating section. They are WR137, WR159, and WR187.

Specification of their inside dimensions is as follows. 

WR137 inside dimensions: 1.372” x 0.622”

WR159 inside dimensions: 1.59” x 0.795”

WR187 inside dimensions: 1.872” x 0.872” 

Maximum rf power transmitted via oxygen-free high conductivity (OFHC) copper WR187 waveguide may be up to 350 MW peak in vacuum \cite{matsumoto1997development}. High-power experiments have shown that the main reason for the rf breakdown is a poor-quality inner waveguide surface.  The waveguide thickness of 0.157” is enough for easy machining and assembling for the brazing process. The 0.157” waveguide thickness may be chosen to allow using a common jig to join each element in the brazing process. Typically, a dimensional accuracy of less $\pm$40 mils is acceptable. 

The theoretical attenuation constant in WR187 is 3.635 mNp/m. It corresponds to the 0.032 dB/m transmission losses at room temperature. A preliminary length of each waveguide branch from the klystron to the cryomodule input is 30~ft (approximately 10 m). An average RF power loss in the copper WR187 waveguide will be less than 1.5 kW in this case.

\item\textbf{RF Flange}

An important issue to address is the design and adoption of a standardized waveguide vacuum flange. This important component must serve in two roles as both an rf seal and a vacuum seal.  The VSWR of the joint must be less than 1.015 at a frequency 5712 $\pm$ 50 MHz. The flange pair  should be capable of transmitting 160 MW peak RF power. A leak rate should be less than  2x10$^{-10}$ std cm$^3$ He/sec. The completed seal should be leak tight when the assembly is subject to a bake-out cycle consisting of heating at rate of 100$^o$C/hr to a peak of 560$^o$C, holding at this temperature for 48 hours and cooling at a rate of 100$^o$C/hr.

Further, the most commonly occurring problems such as discharge breakdown and vacuum leak often happen near the flange gasket. 

A new type unisex waveguide flange has been developed at KEK \cite{matsumoto1997development} to increase reliability and reduce cost; it comes from the RIKEN-DESY WR187  flange as shown in Figure \ref{fig:flange}.

%%%%%%%%%%%%%%%%%%%%%%%%%%%%%%%%%%%%%%%%%%%%%%%%%%%%%%%%%%%%%%%%%%%%%%%%%   
\begin{figure}
\begin{center}
 \includegraphics[trim={0 0cm 0cm 0cm}, clip, width=0.5\hsize]{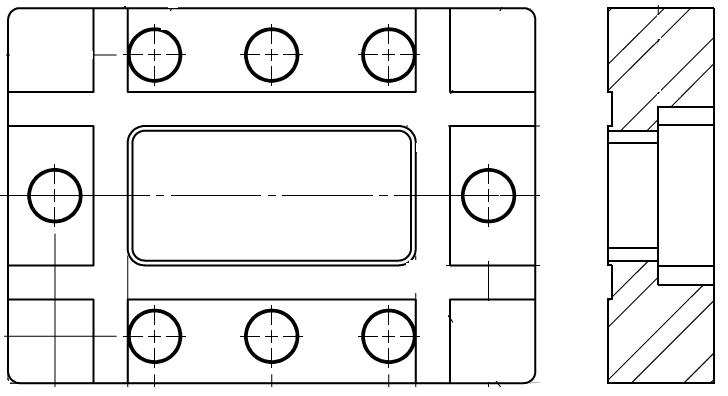}
\end{center}
\caption{Schematic of the face and profile of a RIKEN-DESY WR187  flange.}
\label{fig:flange}
\end{figure}
%%%%%%%%%%%%%%%%%%%%%%%%%%%%%%%%%%%%%%%%%%%%%%%%%%%%%%%%%%%%%%

\item\textbf{RF Window}

The average high-power klystron lifetimes are typically $\sim$50k hours. The klystrons must be replaced in a case of its failure. The replacement work will be performed when a part of the vacuumed rf feeder is opened for disassembly of  the klystron. The rest of the rf feeder will be under the vacuum. The rf window with the ceramic barrier is needed as an interface between an atmosphere and the vacuum envelope.

The rf windows must reliably operate under 80~MW peak power levels. However, in the case full reflection from the accelerating section, the rf window must be working even with double rf power levels of the normal operating mode. A MW peak rf power produces an electrical stress inside the vacuum envelope and ceramic interface. RF breakdowns  in these megawatt power environments could damage the rf window if the design is not done correctly. There are several innovations in the present rf window designs for S- and X- Band frequency ranges. Many of these innovations are practically realised. For example, the X-Band travelling wave mixed-mode rf window \cite{kazakov1998new,yamaguchi1999high,tokumoto2000high} tested in the pulse mode up to 100 MW. The rf window for C-Band will be designed, tested, and optimized under ideas and innovations discussed in \cite{kazakov2007}, Figure \ref{fig:TWwindow}.

%%%%%%%%%%%%%%%%%%%%%%%%%%%%%%%%%%%%%%%%%%%%%%%%%%%%%%%%%%%%%%%%%%%%%%%%%   
\begin{figure}
\begin{center}
 \includegraphics[trim={0 0cm 0cm 0cm}, clip, width=0.75\hsize]{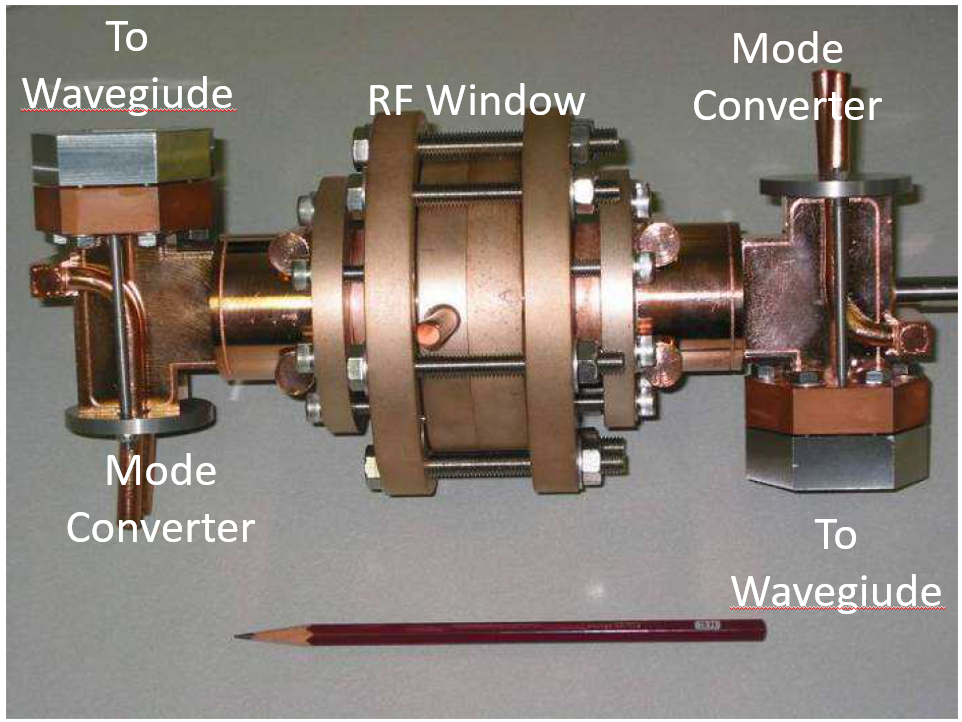}
\end{center}
\caption{X-Band travelling wave mixed-mode rf window\cite{kazakov2007}.}
\label{fig:TWwindow}
\end{figure}
%%%%%%%%%%%%%%%%%%%%%%%%%%%%%%%%%%%%%%%%%%%%%%%%%%%%%%%%%%%%%%

\item\textbf{Power Divider}

One power divider will integrate into the waveguide network to split equally the output klystron power for two branches. A short-slot, narrow wall, coupled junction rf power divider is a tradition rf device that has been used in the high power rf systems. It has high power handling capacity and each of its ports lies in the same plane. The high power tests indicates that a short-slot divider can operate without rf breakdown at more than 70 \% of the power capacity of the terminal waveguides.

The basic construction of the short-slot divider uses two rectangular WR187 waveguides that share a narrow wall. Between them a single slot couples half of the incident power from one waveguide into the second guide. The two resulting waves are 90 degrees out of rf phase. A simplified sketch is shown in Figure \ref{fig:divider}. The electrical phase length matching of both waveguide arms will be performed on the regular straight sections.  

%%%%%%%%%%%%%%%%%%%%%%%%%%%%%%%%%%%%%%%%%%%%%%%%%%%%%%%%%%%%%%%%%%%%%%%%%   
\begin{figure}
\begin{center}
 \includegraphics[trim={0 0cm 0cm 0cm}, clip, width=0.55\hsize]{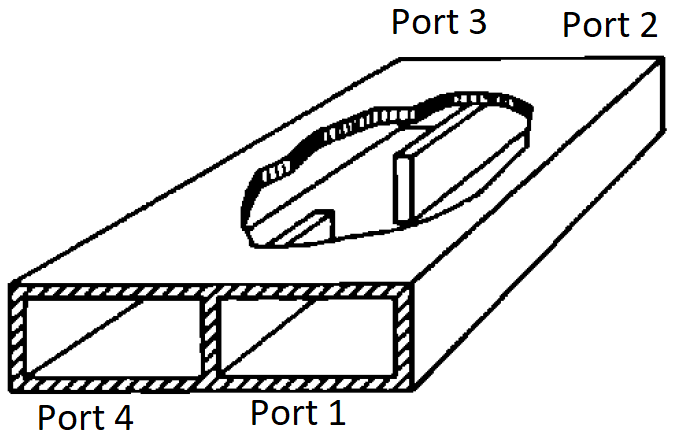}
\end{center}
\caption{Schematic of a short-slot divider using two rectangular waveguides that share a narrow wall.}
\label{fig:divider}
\end{figure}
%%%%%%%%%%%%%%%%%%%%%%%%%%%%%%%%%%%%%%%%%%%%%%%%%%%%%%%%%%%%%%
The \CCC accelerating structures operate in a standing wave. There will be a reflection of power from the structure back to the klystron. Part of the reflection power (ideally all) will propagate into Port 4. The high power vacuum dry rf load will absorb this power.

\item\textbf{RF Load}

There are several concepts for high power load designs. Concepts may be classified by the class of rf absorption material to be employed. The most popular rf absorber that is used in the high power loads is water or other liquid lossy dielectrics. This load concept requires a solid dielectric interface separating the accelerating vacuum envelope and the liquid absorber. Such a concept is not acceptable for the \CCC demo because a catastrophic failure of one interface could destroy the long linac vacuum system; recovery from this failure would be expensive. As a result, the high power load concepts based on the liquid rf absorbers are not being considered. Our analysis will be focused on high power load concepts that are based on a multipactor-free concept with an advanced rf absorption materials capable of working in an UHV environment.

\item\textbf{RF Monitors }

Modified Bethe hole direction couplers are a typical components that have been integrated into the waveguide network for rf power monitoring. The high power level in the waveguide requires the use of more than 60 dB coupling ratios. Three directional couples are shown in the rf feeder in Figure \ref{fig:Network}. This figure represents a typical layout of one klystron station. Many couplers will be needed for the whole installation. A modified Bethe hole directional coupler is optimised to reduce their cost. The designs  used allow for the insertion of a waveguide pump-out and vacuum gauge in the same space where a coaxial strip-line loop is placed. There are two types of designs of the couplers in the SLAC linac. These designs will be evaluated in the \CCC demo. An employment of a  dielectric window on the wide wave guide wall is used in the firs coupler. This dielectric window is placed above the coupling hole and separates a vacuum atmosphere regions. The strip line loop is placed from the atmosphere side of the dielectric. Two vacuum rf feedthroughs are employed in the second concept. Both concepts are a subject for evaluation in the \CCC demo installation.

\item\textbf{Phase Shifter}

One klystron feeds two accelerating sections in the present rf feeder layout. If the two waveguide branches of waveguide network are tuned properly, the rf power from the klystron into the input port of the power divider will arrive in phase at each of the two output ports. There is a transition located a the cryomodule for the waveguide to go into the \LN cryostat where the temperature will drop from room temperature to $]sim$80~K. The temperature decoupling in the waveguide is needed to reduce the thermal losses in the cryogenic vessel. There are several approaches for realizing this thermal break. One of them is to employ the thin-wall waveguide fabricated from a material with low thermal conductivity and low rf losses on the inner waveguide surfaces (for example with a thin copper coating). Inclusion of a phase shifter based on waveguide wall deformations  will be evaluated in this space.

\subsubsection{Start-to-end simulation}
\label{sec:S2E}
High fidelity start-to-end beam dynamics simulations were used in x-ray free electron laser light source accelerator design to verify and optimize the accelerator design concept~\cite{qiang2022,qiang2017,qiang2016,qiang2014}.  
This type of simulation should be carried out for the \CCC accelerator design too.
Important physical effects such as three-dimensional space-charge, longitudinal and transverse wakefields, coherent and incoherent synchrotron radiation, and intrabeam scattering should be included in the simulation model. The effects of long-range wakefields on beam quality should also be studied. Machine imperfections such as misalignment errors and field amplitude and phase errors should be studied using the start-to-end simulation. 

To provide the required luminosity for a linear collider, it is critical to reduce emittance dilution in the main linac from machine tolerances such as cavity misalignments and imperfections~\cite{akcelik2008, lunin2018}. For the \CCC linac, misalignments can arise from individual cavity misalignments and changes in the properties of coupled HOMs in a cryomodule with misaligned and deformed \CCC cavities. The rf parameters and fields of the HOMs can be evaluated for random distributions of cavity offsets (in a cryomodule) and cavity deformations along the full linac. The fields are then used for beam emittance dilution evaluation. A statistical analysis using the constraints form realistic fabrication and component placement tolerances will be facilitated by the HPC capabilities of advanced simulation codes~\cite{bizzozero2021}.

Dark current and capture may have deleterious effects on the particle detector in the form of unwanted backgrounds. Integrated simulations using rf and beam dynamics codes will provide the needed tool for start-to-end simulation for linear colliders. The simulation starts with electrons emitted from cavity walls governed by field emission and their subsequent movements are tracked under the influence of the cavity accelerating mode operating at a specified phase. Secondary emissions will be generated if some of the electrons hit the cavity wall. The exit electrons from the cavity will be used for beam tracking along the linac with magnet components from the lattice until they reach the next cavity. A pipeline workflow involving rf and beam dynamics simulations will be developed to facilitate start-to-end simulations.

\end{itemize}

\subsection{Parallel Research and Development}
\label{sec:parallelR&D}
%\begin{enumerate}
%\item Damping rings
\subsubsection{Damping Rings}

One important issue in advanceing the damping ring R\&D is the injection/extraction system for the train of bunches which are separated a few tens of nanoseconds. Powerful and fast solid-state switches driving a transmission line kicker are needed. They must work in a pulse mode and be capable of generating power from MW to GW and rise/fall time from 10x ps to 100 ns. There are practical physical effects that can overcome the speed limitation for commercially available off self-solid-state switches. For example:

\begin{itemize}

\item{nonlinear ferromagnetic media }

\item{solid-state electron-hole plasma in semiconductors }

\end{itemize}

\begin{figure}
\begin{center}
 \includegraphics[trim={0 0cm 0cm 0cm}, clip, width=0.75\hsize]{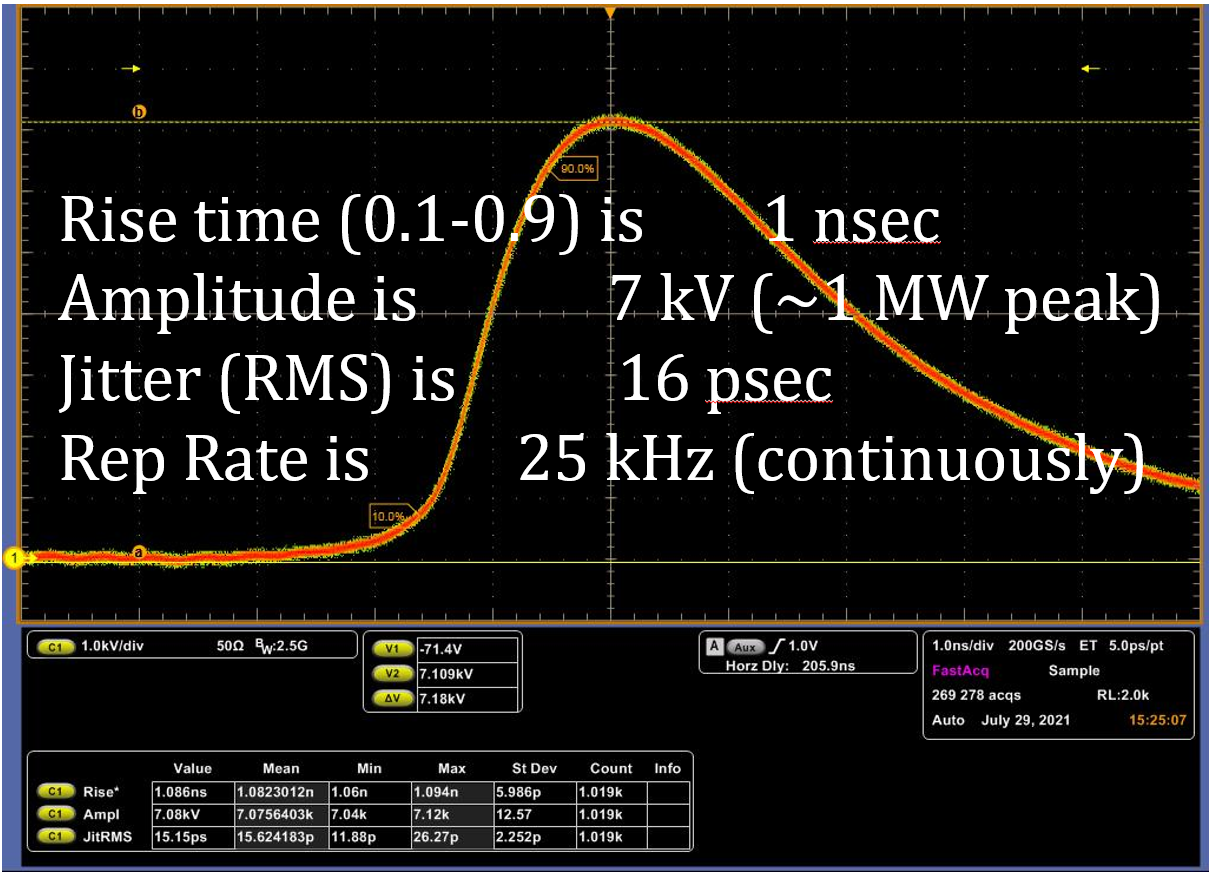}
\end{center}
\caption{Demonstration of a 1 MW/ns switching speed.}
\label{fig:Pulser}
\end{figure}

 In both approaches the nonlinear characteristics of the switching media (ferromagnetic and semiconductor) are employed in the final stages of the pulser to form multi- MW level nanosecond pulses. The pulse dynamic processes in ferromagnetic and semiconducting materials can support a 1~MW/ns switching speed stably, reliable, and efficiently. This fact is confirmed by our pilot experiments. 
For example, the effect of the high-power switching speed in semiconductors diodes is shown in Figure \ref{fig:Pulser}. 

In this case diodes were fabricated with special doping gradients in Si-based wafers via a deep diffusion process.

The waveform shown in Figure \ref{fig:Pulser} demonstrates a 1.3 MW/ns switching speed continually running at 25 kHz repetition rate with less than 20 ps RMS time jitter. A  development of physics basis of semiconductor behaviour for a generation of GW levels of nano- and sub-nano power swings would be useful for the development of the damping ring injection/extraction system in the \CCC machine.  

We will explore a variety of ultra-wide band-gap (UWBG) semiconductors (diamond, SiC, \textit{etc.}) which are of great interest for their efficiency, speed, thermal management and radiation resistance. 

%%%%%%%%%%%%%%%%%%%%%%%%%%%%%%%%%%%%%%%%%%%%%%%%%%%%%%%%%%%%%%%%%%%%%%%%%   
%%%%%%%%%%%%%%%%%%%%%%%%%%%%%%%%%%%%%%%%%%%%%%%%%%%%%%%%%%%%%%

%\item Beam Delivery System and Final Focus (BDS/FF)
\subsubsection{Beam Delivery System and Final Focus (BDS/FF)}
\label{sec:BDS}

Mature designs exist for beam delivery systems with an electron or positron beam energy in the range of 125 GeV – 1.5 TeV as part of the ILC and CLIC programs. \CCC will continue to leverage the significant effort in simulations and hardware tests that have been developed and carried out over the past 30 years in preparation for a future linear collider. Trade-offs exist in design length and aggressiveness of beam parameters (e.g. $\beta$*,L*). These manifest as tighter, more complex diagnostics/correction hardware (higher-order magnets, $<\mu m$ accuracy movers, BPMs etc) and more complex, slower online beam size tuning systems, limiting achievable luminosity. The design for the BDS and FF is complicated and time consuming to evaluate. To date, a final focus system with ILC levels of demagnification has been experimentally demonstrated \cite{ATF2}, but not beyond. We will use caution when designing the BDS and FF for \CCC with the main goal of simplifying and reducing the length of the BDS and FF. Pushing the design beyond key ILC FF parameters (principally, energy spread and final focusing doublet chromaticity $\sim L^*/\beta^*$) would require further experimental verification of performance.

The backbone for assessing the realistic luminosity potential of any linear collider is that of the start-to-end simulation (S2E). This should include a full description of the beam transport from the exit of the source (damping ring or rf photo-injector) through to the BDS and FF and to the collisions at the interaction point (IP). All relevant physics processes should be included: e.g., chromatic effects of magnets, short and long-range longitudinal and transverse wakefield kicks in accelerating structures, synchrotron radiation in bends, magnetic field errors and placement errors of all accelerator components. Both “static” misalignments and other errors as well as dynamic effects (errors which are fast on the timescale of corrective actions being applied) need to be considered. The BDS is especially sensitive to errors, containing diagnostics systems required by the physics detectors experimental program, and strongly non-linear beam dynamics with large multi-km beta functions, for example. Careful simulation with accurate descriptions of the various errors internally and input conditions from the linac are critical to assess the final performance of the collider. Extensive S2E simulation programs have been implemented in the past for the ILC and CLIC communities, with a strong focus on the FF tuning system \cite{FFTUNE}. It is envisioned to re-create such a community of beam dynamics simulation experts to generate a S2E simulation environment to assess the \CCC performance and form the basis for tolerance specifications.

The use of Machine Learning (ML) and Artifical Intelligence algorithms (AI) in the design and operation of particle accelerators is somewhat new and has not yet been fully explored for the field of next-generation linear colliders. The \CCC program should benefit from this rapidly advancing area of research. The BDS and FF are strongly non-linear systems which require complex tuning procedures to achieve full luminosity \cite{FFTUNE}: Application of modern, multi-objective physics-informed Bayesian optimizers could yield powerful improvements in tuning performance and speed, possibly increasing confidence in an ability to push on the FF parameters. This would manifest in either an increased luminosity performance or operation at a similar luminosity with reduced beam power requirement (cost). Another example application of ML/AI technology could be applied to the problem of reconstructing beam aberrations at the IP collision using a combination of ``event shapes" reconstructed using ``beamstrahlung" monitors situated close to the IP \cite{ipparreco}. Again, such technology could help with the accuracy of the beam tuning process yielding increased tuning speed, whilst also providing useful information to the detector community.

As exhaustive experimental tests have shown, at ATF2~\cite{ATF2}, and FFTB~\cite{FFTB} before it, despite the highly non-linear design aspects and complex tuning procedures, modern optics design and beam dynamics software is well capable of accurately describing the BDS and FF systems. Whilst there is always value added in test facilities, a test final focus system demonstration for \CCC is likely unnecessary as it does not strongly deviate from existing designs for the BDS/FF. A possible exception is the particular case of the FF design for the $\gamma$-$\gamma$ design (XCC)~\cite{XCC}. This option requires the focusing of a round-beam configuration, necessitating a final triplet configuration of magnets which has not yet been designed or studied for small-beta focusing optics required at a linear collider. One possibility to demonstrate this configuration for the FF might be a future upgrade to the FACET-II~\cite{facet2} Sector 20 experimental region which already operates with round-beams (E=10 GeV, $\beta$*$>$5cm, s$_{z}$ $<$1um, Q$<$2 nC). Further novel aspects of the XCC concept which demand a fresh look at the BDS design include the integration of keV energy x-ray focusing optics to overlap with the electron beams close to the IP and generate the colliding gamma beams.
Consideration of FF optics which pushes substantially beyond that already demonstrated can be beneficial in terms of cost savings and/or luminosity performance. The premier facility in the world for such experimental studies is ATF2 at KEK: whilst the accelerator is currently still operating, serious further investment would be required to enable a program to experimentally verify such pushed parameters are feasible.

\subsubsection{Levitated Positron Target - Radiatively Cooled}

Positron sources are based on electron or photon beam striking a high-Z target, often Ti. Because both the instantaneous and average power are high, the concept of a rotating Ti hoop in the accelerator vacuum is accepted. The the ILC baseline, water cooling introduced through rotating seals, and rotating seals are used for an axial drive shaft. A concept has been proposed\cite{breidenbach2016positron} using radiation cooling of high performance Ti alloy blades to a water cooled vacuum can, completely eliminating water channels in the vacuum space has been proposed. In addition, a concept for magnetic suspension and drive of the hoop,  eliminating shaft seals, that does not require permanent magnets, and so should survive in the high radiation field of the target is also being explored. Increasing the reliability and performance of the positron target would have significant benefits for \CCC operation and advancing this concept will be explored.

%\item Levitated Positron Target - Radiatively cooled 
%\item QD0/QD1 
%\item Advanced RF Source Research and Development
\subsubsection{Advanced RF Source Research and Development}

With the development of this highly efficient linac topology, we expanded our vision for a future high gradient linac architecture that comprises a much more simplified RF system with a great deal of optimization for the cost of the peak power generation. A central tenant of the \CCC concept is a rf only upgrade to increase the energy from 250 GeV to 550 GeV CoM. This upgrade affords us the opportunity to implement novel concepts for the rf sources that will be used for the upgrade. To this end, we are proposing a very simple, modular system with an order of magnitude cost reduction for the accelerator complex. The idea starts with developing low voltage RF sources with extremely high efficiency, which translates to a low current and lower power. To get to high power, we utilize the narrow bandwidth of the system, which allows us to design an efficient combining network for a large number of low voltage systems. This, in turn, leads to inexpensive, relatively low voltage modulators and hence fast rise and fall times,  eliminating the need for pulse compression. With a large number of small tubes, mass production cost reduction could result in substantial savings. 

To realize this vision, we need to have a comprehensive program for these advanced microwave rf sources\cite{weatherford2020modular} with the end goal of an experimental demonstration of C-band modular multi-beam klystrons with efficiencies $\geq$ 65$\%$, with periodic permanent magnet (PPM) focusing and high power. This should be followed by a demonstration of the combining mechanism; many methodologies are possible, but there is a new invention for combining with a two-dimensional Flouquet network; a 16 tube combiner has already been demonstrated\cite{maxim2019phaser}. Finally, a demonstration of a fully integrated tube with suitable power combination operating with efficiency $\geq$65$\%$ should follow. 

\subsubsection{RF Pulse Compression}

Another possible upgrade to the \CCC concept is the implementation of rf pulse compression. RF pulse compression can decrease the fill time and thermal load into the cryogenic vessel. Given the efficiency of cooling at cryogenic temperatures of $\sim$80~K (approximately 15\%) this can have a significant impact on the cost and operational efficiency for the complex. A compression of 3X for the filling portion of the rf pulse has been shown to have tremendous benefits\cite{bane2018advanced}. The development of pulse compressors for cryogenic linacs that meet this level of compression is an ongoing topic of R\&D, as noted in the discussion of Very High Energy Electron therapy systems in a concurrent submission to Snowmass 2021 on FLASH Radiation Therapy \cite{FLASH}. RF pulse compression has been used extensively for large scale accelerator applications to supply short high peak power pulses with cost efficient long pulse, low power sources\cite{farkas1986binary}. Recent innovations in RF compressor cavity design have enabled a dramatic reduction in the system footprint while maintaining the capability to produce 6-fold pulse compression and isolate the source from the reflected RF signal from the cavities.\cite{wang2017development,franzi2016compact} For cryogenic copper structures, cavity designs will need to achieve high intrinsic quality factors, Q$_0$ up to 400,000, and high coupling factors, $\beta$ up to 10. Active research in this area will continue to benefit the broader accelerator community that relies on pulse compressors to supply the peak powers needed for high gradient operation. Indeed, the development of accelerator technology needed for \CCC, from the distributed-coupling linac design at cryogenic temperatures to potential pulse compression for high-efficiency operation, will have a cross-cutting impact in support of efforts to design compact, cost-efficient accelerators for medical and security applications.  

%\item RF Distribution (pulse compressor)

%\item Error Sensitivity Study 
%\item Polarimetry 
%\item Coatings 
%\end{enumerate}

\subsection{Industrialization}

%\begin{enumerate}
%\item Cryogenics Quads 
%\item Linac Fabrication

As with any linear collider designs the volume of hardware production that will be required is a daunting task for \CCC. In an effort to render the required production practical we will strive to engage with the existing and emerging accelerator technology sector and related technological sectors that will be required for \CCC. Fortunately, the production of hardware directly applicable to a normal conducting rf accelerator is already quite significant. This is in large part driven by the demand for medical and industrial accelerators that produce high-flux x-rays. For example, in less than a decade from the  launch of Varian's flagship radiation therapy model, one company built and installed the equivalent of a \CCC main linac in hospitals all over the world.\cite{Varian} If \CCC technology is successfully transitioned to the industrial sector not only will \CCC benefit from industrial production, but commercial vendors will be able to offer more compact and higher efficiency accelerator and rf sources. 

This potential exists beyond the direct accelerator technology sector. For example, one of the key advantages of the \CCC accelerating structure is the minimization of part count, a robustness of manufacturing through CNC end milling of split block structures, and use of bonding techniques that minimize the possible impacts on structure performance. While commercial vendors are presently integrated in the production of linac components, increasing the rate of fabrication, reducing part-to-part variability, automating tuning and diversifying the supply chain are all topics to be explored during the demonstrator R\&D plan. The production of accelerating structures for three cryomodules, individual structure tests and the injector, will allow for a broader engagement with manufacturing industry (in particular small-business machine shops with aerospace experience).

The RF sources (klystrons) are expected to be significant cost drivers for any future accelerator facility, and \CCC is no different in this regard. To drive down klystron procurement costs, commercial partners should be engaged as soon as possible in the development effort. There should be robust support for a public-private collaboration to evaluate the most cost-effective path forward. A few options could be considered, depending on the available infrastructure, tradeoff between lab-run or supplier-run operations, and the capabilities of industrial partners:
\begin{itemize}
    \item Klystron design, fabrication, and test entirely by multiple commercial partners. It is important to ensure simple interchangeability of sources in this situation.
    \item Klystron design completed by government labs; klystrons built-to-print, tested, installed, and maintained by commercial partners.
    \item Klystron design completed by government labs; built-to-print by commercial partners; but tested, installed, and maintained by the \CCC{} facility.
\end{itemize}
Industrial R\&D is driven by immediate and long-term market potential. From this viewpoint, a proposed future collider presents a significant risk that may not merit substantial investment in cost reduction and process improvements by industry.  Therefore, in any partnership scenario, the industrialization effort should require – and financially support – a collaboration with government labs to evaluate and improve upon manufacturing processes. When the most economical approach is to use existing fabrication methods, those methods should be leveraged. When process improvements can be made (for example, using additive manufacturing to reduce part counts, or automation of testing processes), the investment required for industry to implement these changes should be supported directly. This provides a “win-win” scenario where the klystron costs for the future \CCC klystrons are reduced, and commercial partners can benefit from cost reduction measures which are broadly applicable and attractive to industry participants regardless of the timetable for the \CCC machine itself.

With this in mind, during the \CCC Demonstration R\&D timeline (and in many cases presently ongoing) we will directly engage with industry in many technological areas: accelerator fabrication, beamline components, permanent magnets, vacuum vessel production, rf sources (klystron and modulator), LLRF, tunnelling, large scale cryogenics, \textit{etc.}

%\item Cryomodule Cryogenics and Beamline Design 
%\item Cryomodule Production 
%\item Klystron 
%\item Modulator 
%\item Tunnel 
%\item Pre-Fab Surface Construction 
%\item Large scale cryogenics 
%\end{enumerate}

\subsection{Additional HEP Research and Development}
\label{sec:HEPRD}
%\begin{enumerate}
%\item AI/ML

\subsubsection{Artificial Intelligence and Machine Learning (AI/ML)}

%\begin{comment}
Modern Artificial Intelligence (AI) and Machine Learning (ML) techniques can be utilized with great efficacy during  accelerator design in the CDR/TDR efforts and for machine control during operation of the collider. The demonstration accelerator also provides interesting opportunities for co-design of components and methods for ML-based modeling and control.
%open source tools for running design optimization + incorporating proven ML-based techniques already available

During the design work, a number of accelerator parameters that are correlated need to be studied, and optimized combinations of parameters will need to be found prior to performing detailed simulations. We plan to use and potentially expand upon ML-based optimization techniques that can identify the most important combinations of parameters and parameter ranges that need to be explored in detail for optimal setups, likely resulting in a faster design process, reduced burden on computational and personnel resources, and superior optimization results.ML-based optimization methods are able to perform high-dimensional multi-objective optimization efficiently \cite{roussel2021multiobj}, allowing for many tradeoffs between output beam parameters of interest to be examined simultaneously. Bayesian methods aimed at automatically exploring accelerator parameter spaces in a sample-efficient way have also been tested in simulation \cite{roussel2021turnkey} and experimentally. These have been incorporated into open-source software tools for accelerator simulations and optimization that are readily available for use in the design process \cite{christopher_mayes_2021_5559141,lume}. 

These methods can also be used for characterizing and optimizing the accelerator during startup and commissioning, and we plan to integrate ML heavily into the commissioning process. Bayesian exploration and optimization can be used without extensive previous data, and can be conducted in a way that incorporates learning to avoid constraint violations (such as a beam going off of a screen, or a beam parameter moving too far away from an optimal band) \cite{roussel2021turnkey}. This makes these methods ideal for optimizing the accelerator incrementally during early commissioning. Information from physics simulations generated during the design process can also be used to help inform Bayesian optimization and speed up its convergence \cite{hanuka2021physics,duris2020bayesian}. Data gathered during commissioning can then be used for automatic calibration of physics models to better match machine measurements and potentially identify early sources of systematic mis-match between the two that can be corrected either on the machine or in the physics simulation \cite{Ivanov2020}, as well as adapted over time to make model predictions more robust to time-varying sources of error \cite{scheinker2021adaptiveML}. During regular operation, ML-based models can be used for making fast online predictions of beam behavior and fast feed-forward corrections (e.g. see  \cite{edelen2016neural,scheinker2018demonstration,azzopardi2019operational,Leeman2019demonstration}). Feed-forward corrections can potentially be used to improve beam stability and incorporated into standard continuous control tasks (such as trajectory control and low level rf control) for improved performance. 

 There is also a relatively unprecedented opportunity for co-design of the accelerator diagnostics and control methods. The placement and types of diagnostics can be optimized to maximize useful information from the machine and reduce redundancies. This could be paired with running simulations of the anticipated online modeling and control techniques that will be used once the machine is built. For example, several ML approaches have been demonstrated for inferring unseen aspects of the beam behavior based on a subset of available measurements (e.g. \cite{emma2018machine,scheinker2021adaptiveML,VD_spectral}). The  expected performance of these methods could be taken into account  directly by integrating simulated versions of them in the design of diagnostics placements. Several other examples of possible applications of AI/ML are discussed in Sections \ref{sec:S2E} and \ref{sec:BDS}.
%\end{comment}
%online predictions/feedforward control

%co-design

%will reincorporate text into the above
%ML methods can be used to detect abnormal operation scenarios 

%\item Plasma Lens 
\subsubsection{Plasma Lens}

Plasma lenses can focus electron beams with strengths several orders of magnitude stronger than quadrupole magnets~\cite{Doss2019,chen:1989prd,chen:1990prl}. In passive plasma lenses, the transverse focusing force in the underdense, nonlinear blowout plasma wake regime is due to the presence of the stationary plasma ions. If the transverse density profile of this ion column is uniform, then the focusing force experienced by the electrons in a relativistic beam is both axisymmetric and linear. These properties lead to an aberration-free focus of the beam that can achieve unprecedented small beam spots. The first order beam dynamics have been described in Ref.~\cite{Doss2019}.

The underdense plasma lens (UPL) operates in a two-bunch configuration where a ``driver'' electron bunch drives the nonlinear wake and a second, ``witness'' electron bunch is subsequently focused in the nonlinear blowout wake.  It may also be possible to operate in a single-bunch configuration where the head of the electron bunch drives the nonlinear wake and the bulk of the electron bunch is focused.  In this regime, the linear focusing force from the ions results in a focal length given by the focusing strength
\begin{equation}\label{eqn:uplf}
f\equiv\frac{1}{KL}=\frac{1}{2 \pi r_e}\frac{\gamma_b}{n_p L}.
\end{equation}
Here, $K$ is the focusing strength, $L$ is the lens's longitudinal thickness, $r_e$ is the classical electron radius, $\gamma_b$ is the Lorentz factor of the electron beam, and $n_p$ is the plasma density of the lens.  A UPL with a plasma density of $n_p=1\times 10^{17}\mathrm{\ cm^{-3}}$ can have a focusing strength $K=88400\mathrm{\ m^{-2}}$ and focal length $f=3.3\mathrm{\ cm}$.  For comparison, a conventional quadrupole with field gradient $G=1\mathrm{\ T/m}$ would have $K=0.3\mathrm{\ m^{-2}}$ and focal length $f=1000\mathrm{\ cm}$ and a permanent magnetic quadrupole with $G=500\mathrm{\ T/m}$ can expect $K=150\mathrm{\ m^{-2}}$ and focal length $f=81\mathrm{\ cm}$.

The UPL can be generated by laser ionization of gas, e.g. by focusing a femtosecond laser pulse into a gas jet. To minimize the footprint, the laser pulse can propagate transverse to the electron beam axis, reducing the required space along the beam line to millimeters.  The longitudinal density profile of the plasma lens is then given by the laser parameters and focusing optics for gas ionization.  Something as simple as a Gaussian focus from a spherical lens with a variable offset from the laser focus to the beamline can give a plasma lens thickness $10's - 100's\mathrm{\ \mu m}$.

There has been discussion about the potential use of a plasma lens for the final focusing element of a collider for decades~\cite{chen:1989prd,chen:1990prl}. The strong, linear, axisymmetric focusing of the UPL in combination with its ultra-compactness and ``self-aligning'' characteristics may show it to be a viable candidate, although further study is still required. It should be noted that the functionality of the UPL described above applies only to negatively charged relativistic leptons. Relativistic positrons would experience a linear {\it defocusing} force in this scheme. It may be possible to achieve similarly strong focusing with the UPL if the positrons are sent through the lens in a different phase of the nonlinear wake, but this is speculative. Another important consideration is the scattering of the beam off of the plasma ions, which may produce both an increase in the beam emittance and a forward-directed shower of secondary particles. The former could result in a reduced luminosity and the latter could result in an increased background for the particle detectors.

The possible application of such a technology to \CCC would require experimental investigation with the bunch structure and beam format of \CCC. Experiments targeted at understanding this technology and associated issues such as plasma recovery time \cite{d2022recovery} could utilize the \CCC demonstrator facility.

%\item Muon Production
%\item PWFA 
%\subsubsection{PWFA}

%\item XCC
\subsubsection{XCC}
\label{sec:XCC}
The XFEL Compton Collider (XCC)\cite{XCC} is an X-ray free electron laser (XFEL) based $\gamma\gamma$ collider Higgs factory concept in which
  $62.8$~GeV electron beams collide with  1~keV X-ray beams to produce colliding beams of 62.5~GeV photons. The Higgs boson production rate is 
32,000 Higgs bosons per $10^7$ second year, roughly the same the ILC rate. The XCC  takes advantage of
\CCC technology in several ways, and its development could in turn help further  \CCC technology and the \CCC $\ee$ collider.

The use of \CCC technology in the XCC starts with the injector, which has the same specification as the
cryogenic RF high-brightness polarized photo-injector described in~\ref{sec:HBpolarizedgun}, with the exception that asymmetric emittances are not needed for $\gamma\gamma$ collisions.
The XCC linac is based on \CCC technology, operating at a bunch train rate of 240 Hz, with 76 bunches per train in the first half of the 63~GeV linac and
38 bunches per train in the 2nd half.  With a linac gradient of 70~MeV/m, the XCC footprint is about 2.5 km.
Given the relatively short linac length and the absence of damping rings, the XCC may provide significant cost savings over $\ee$ Higgs factories.

The XCC XFEL has a pulse energy of 0.7~J, which is 300 times  the pulse energy of current soft X-ray FEL's. The production and focusing of such an X-ray beam
are the greatest technical challenges for the XCC.   These challenges can be addressed by building a 1~nC 120~nm-rad cryogenic RF injector for LCLS-I.  The construction of such an injector would provide the  RF high brightness photo injector demonstrator discussed in~\ref{sec:HBgun}.  When this injector is incorporated into LCLS-I, the upgraded XFEL could be used to investigate the production and focusing of 100~mJ per pulse soft X-ray beams and,  simultaneously, open up exciting new research opportunities in photon science.  The photon science applications of the injector could shift the time frame of its development and construction forward.

\subsubsection{Polarized Positrons}

In the final collider scheme, the availability of polarization for the positron beams can be very beneficial to enhance particular channels, reduce background and facilitate searches for BSM chiral new physics.
The current scheme to generate polarized positrons in the ILC utilizes a very long short period superconducting undulator to obtain a high flux of circularly polarized gamma-rays that can be converted in e$^-$e$^+$ polarized pairs in a thin Ti-alloy target. The scheme assumes the availability of very high energy (125 GeV or more) e-beam and creates a significant dependency burden on the collider complex. Alternative schemes based on Inverse Compton Scattering have been proposed to generate the gamma-rays. The main advantages are the stand-alone nature of the source (since only GeV-level e-beam energy is required) and the possibility by increasing the energy of the gamma-rays to achieve higher polarization purity ($>$ 60 \%) for the e$^+$ beam. One of the major challenges is in the low interaction cross section and related high required average power for the laser, but efficient energy extraction from high quality electron beams could be used in an optical cavity to replenish the laser losses and sustain high repetition rate interactions. 
One important aspect to consider here is the particular \CCC time-format which 
might create a challenge for the cavity thermal management and the heat dissipation on the conversion target. Conversely, the short bunch-to-bunch separation matches well the characteristics of novel laser stacking cavity developments. All of these aspects could/should be tested and evaluated in a demonstrator facility.

%\item cryogun high brightness applications

%\end{enumerate}

%\subsection{Follow on Facility Research and Development}
%\begin{enumerate}
%\item PWFA 
%\item Beam dynamics, extreme bunch compression 
%\item FEL 
%\item gamma ray sources or ICS
%\end{enumerate}

\section{Infrastructure Requirements for Demonstrator R$\&$D}
\subsection{Basic Requirements}
\label{sec:demo_system_basic_req}
Full implementation of the \CCC Demonstrator R$\&$D program has the following facility requirements:
\begin{itemize}
    \item  a radiation shielded bunker at least 50 m long, 3 m wide, and 3 m high. 
    \item a properly shielded dump for a 20 kW 2 GeV electron beam. 
    \item 480 VAC 3 phase power for 21 modulators at 15 kW each, or $\sim$300 kW
    \item Low Conductivity Water sufficient to handle $\sim$300 kW
    \item the site must be suitable for a liquid nitrogen storage facility of at least 25,000 liters (preferably 50,000 liters), with access for liquid nitrogen tanker trucks. 
    \item it must have the electrical, cryogenic, and radiation safety organization for a facility of this scale.

\end{itemize}

\subsection{Siting Options}
\label{sec:siting}
%\textcolor{red}{Discuss in context of what is available?}

Options for locating partial or full-fledged demo facility are:
%\begin{itemize}
%\item \textbf{ESB at SLAC}
\subsubsection{ESB at SLAC}
The Demo Facility could be located in SLAC's End Station B (ESB), where an appropriate shielded bunker is available. This bunker currently houses the NLCTA, which would be removed. ESB has adequate power for the linac RF system. This approach also provides the path for future PWFA based collider R\&D such as staging.

A \CCC Demonstration Facility is envisioned to be located inside SLAC’s End Station B (ESB), where a shielded bunker is available. This bunker currently houses the NLCTA, which would be removed. With the removal of the NLCTA, sufficient electrical power and LCW would be available for the \CCC demonstration facility, see Table \ref{tab:NLCTAparam}. ESB is in SLAC south research yard with heavy equipment and semi-truck access, allowing for delivery and storage of LN$_2$.

\begin {table}[h]
\begin{center}
\begin{tabular}{c c c }
     
\hline
3$\phi$ electrical &	480 V	 & 1200 A \\
\hline
%45$^o$C LCW &	Flow Rate &	Pressure \\
%\hline
30$^o$C LCW &	325 gpm &	85 psig \\

\end{tabular}
\end{center}
\caption {Electrical Power and LCW cooling capacity}
\label{tab:NLCTAparam}
\end{table}

ESB has an operational overhead bridge crane with a maximum lifting capacity of 50 tons, which will be invaluable to the \CCC demonstration facility construction and operation.

The radiation enclosure that houses the NLCTA has the required safety system for machine operation, these critical safety systems include: Personnel Protection System (PPS), Machine Protection System (MPS), Beam Containment System (BCS) and Laser Safety System (LSS). These safety system are supported by SLAC Accelerator Directorate. The NLCTA is activity supporting several experimental programs utilizing the exciting S- and X-Band RF power sources along with a UV photocathode drive laser system. It is envisioned that the S-Band modulator/klystron and UV photocathode laser would be retained and used for the \CCC demonstration facility photo-injector.
To accommodate the 3 \CCC cryo modules, the NLCTA bunker is 51 meters long with an internal height of 3.05 meters. The interior of the NLCTA enclosure has the following dimensions: 51 m x 2.75 m x 3.05 m (L,W,H).  It should be noted that the NLCTA radiation shielding and terminating beam stop were designed for up to 1170 MeV. Minor modifications to the NLCTA West entrance will be required to allow for the installation and maintenance of the cryo rafts/modules.

\begin{figure}[!ht]
\begin{center}
 \includegraphics[trim={0 0cm 0cm 0cm}, clip, width=0.95\hsize]{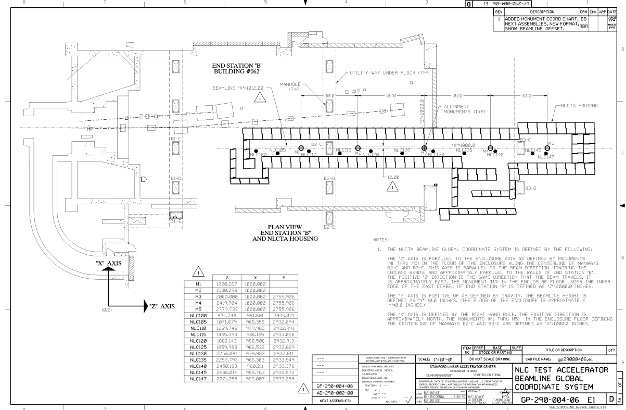}
 \end{center}
\caption{Top-down view of End Station B along with the NLCTA. }
\label{fig:NLCTA}
\end{figure}

%\item \textbf{FAST at Fermilab}
\subsubsection{FAST at Fermilab}
IOTA/FAST is an R\&D Facility for Accelerator Science and Technology at Fermilab. It has two components: an Integrable Optics Test Accelerator (IOTA), 150 MeV electron / 2.5 MeV proton storage ring, \cite{IOTA} with a dedicated proton injector and an SRF linac. The 300-MeV FAST linac serves as an injector of electrons for IOTA and also provides beam to dedicated experiments. 

Originally, the ILC-style SRF linac was envisioned as a demonstration facility to test and operate a full ILC ``RF unit'' with ``ILC beam intensity.'' The RF unit consists of 2 cryomodules driven by a single 10~MW klystron. However, only one cryomodule was installed at FAST and has successfully demonstrated the ILC specification on accelerating gradient of 31.5 MeV/m \cite{broemmelsiek2018record}. The ILC beam intensity is a $\sim$1 ms long train of $\sim$3,000 bunches (3~MHz bunch repetition frequency) with a charge of 3.2~nC per bunch. The bunch train repetition rate is 5~Hz, and the r.m.s. bunch length is 300~$\mu$m.

Beside the 8-cavity SRF cryomodule, the electron injector includes a 5-MeV RF photo-injector of a DESY/PITZ design, a 25-m long low-energy ($\leq 50$~MeV) beamline with two SRF capture cavities, and a $\sim$100-m long high-energy ($\leq 300$~MeV) beamline. Both beamlines are equipped with a suite of high-precision beam instrumentation.

\begin{figure}[!ht]
\begin{center}
 \includegraphics[width=0.65\textwidth]{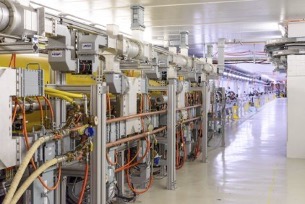} 
\end{center}
\caption{Photo of the FAST high-energy beamline at Fermilab. }
\label{fig:FASTHEbeamline}
\end{figure}

The FAST high-energy beamline (Fig.~\ref{fig:FASTHEbeamline}) has plenty of space to accommodate \CCC demonstration cryomodules and associated high-power rf equipment. The facility has a dedicated cryogenic system, which includes a 5,800~gallons ($> 26,000$~liters) capacity LN$_2$ tank. At first, FAST can be used for cryogenic RF testing of \CCC cryomodules with and without beam. Though the present FAST injector cannot provide beam in the \CCC beam format, a relatively straightforward upgrade with S-band injector can be undertaken for a full-scale beam demonstration.

%\item \textbf{UCLA Research and Infrastructure}
\subsubsection{UCLA Research and Infrastructure}
\label{sec:UCLA}
 
\begin{figure}[!ht]
\begin{center}
 \includegraphics[trim={0 0cm 0cm 0cm}, clip, width=0.95\hsize]{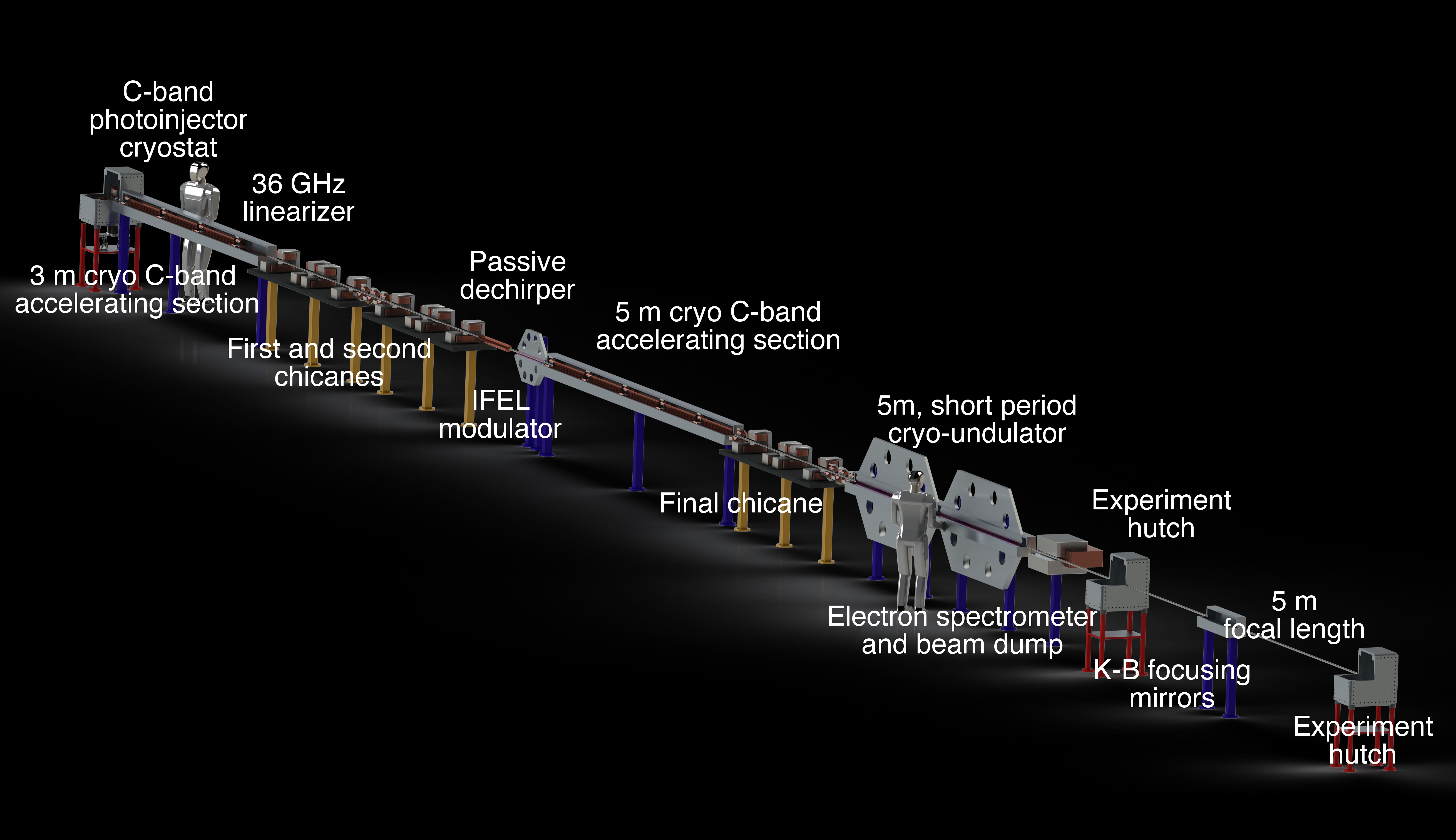} 
\end{center}
\caption{Schematic of UC-XFEL project spearheaded by UCLA. }
\label{fig:UCXFEL}
\end{figure}

UCLA is proceeding with a vigorous program in C-band cryogenic RF structures, with the twin goals of developing a linear collider-capable injector and the first steps towards the realization of an ultra-compact XFEL\cite{rosenzweig2020ultra}. One of these steps is the construction of injector with unprecedented brightness, as in the asymmetric emittance beam case mentioned above. This source has been evaluated, and is expected to be able to drive a soft-x-ray FEL into saturation in 4 meters of very short period undulator. The use of a short period undulator is uniquely enabled by the very high brightness electron beam obtained from the same high-field cryogenic RF photo-injector that produces the linear collider-oriented magnetized beam. 

With a short period undulator one may operated the beam at much lower energy, that is 1 GeV in soft-X-ray case. By using cryogenic C-band linac sections, with C$^3$ design, this energy is achieved in 8 m of active length with 125 MeV/m gradient. As such, the overlap between the  C$^3$ and ultra-compact X-ray FEL (UC-XFEL) initiatives is considerable. The layout envisioned for UC-XFEL is shown in \ref{fig:UCXFEL}. This shows a compact XFEL system, including two-stage compression and compact X-ray optics all located within a 40 m footprint, as shown in Figure \ref{fig:UCXFEL}. The estimated cost of the first prototype of this system is 32M dollars.

The UC-XFEL overlap with C$^3$ testing needs is considerable. An ultra-high brightness beam is available to test linear collider beam dynamics including short-range BBU in a significant stretch of linac. The photoijector itself serves as a test-bed for the linear collider electron source.

\begin{figure}[!ht]
\begin{center}
 \includegraphics[trim={0 0cm 0cm 0cm}, clip, width=0.95\hsize]{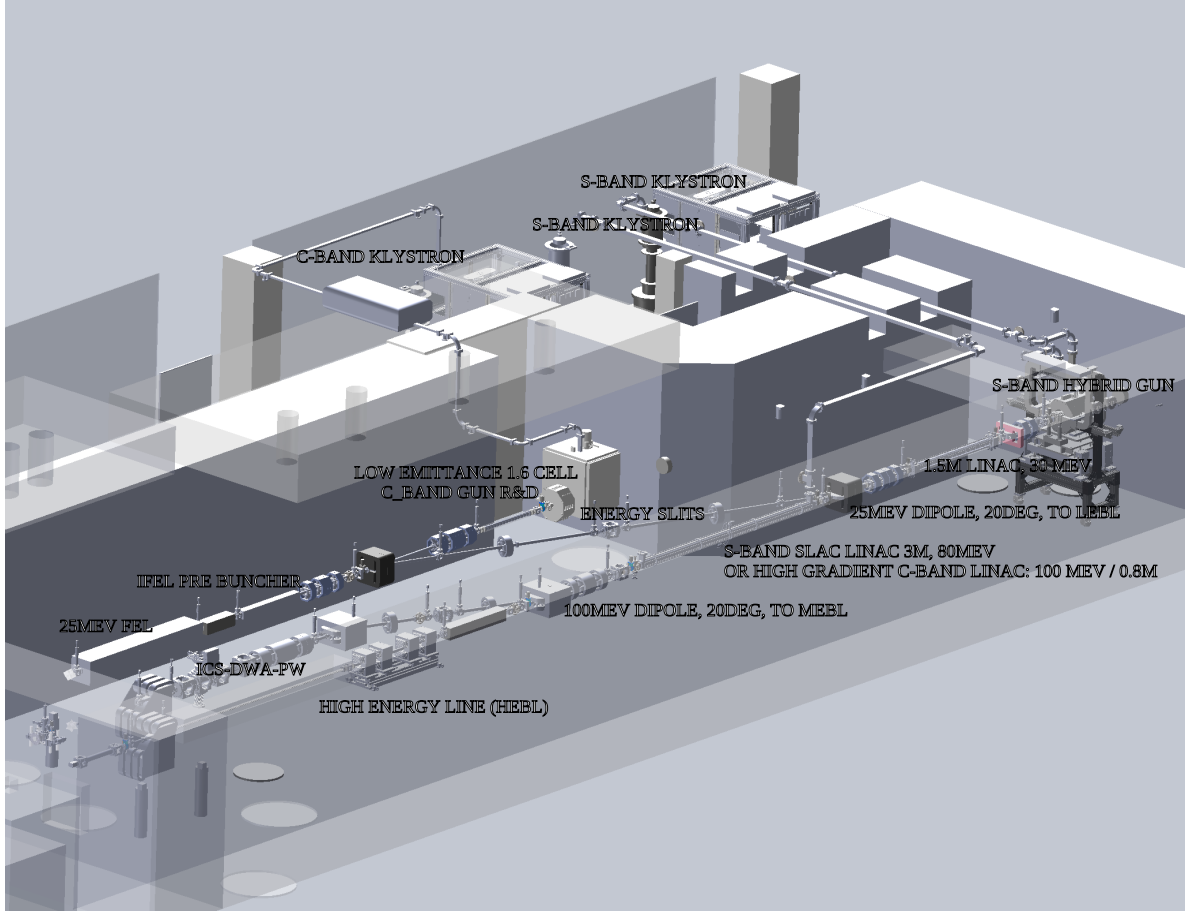}
\end{center}
\caption{Schematic of UCLA SAMURAI infrastructure as currently under development.}
\label{fig:SAMURAI}
\end{figure}

As noted, at UCLA, the C-band cryogenic photo-injector is already under development at the UCLA MOTHRA laboratory, which also hosts a 0.5 cell cryogenic photo-injector having a load-lock system which permits testing of a variety of photocathodes at cryogenic temperatures. This system is being commissioned now. These cryogenic C-band structures are powered by a C-band klystron (5 MW, with a SLED system under development to reach 20 MW). After initial testing, the high brightness photo-injector will be transferred to the UCLA SAMURAI lab,

UCLA SAMURAI is a large (55 ft long) bunker and associated laser room and RF power complex. It is currently being commissioned with S-band accelerator components, with the goal of converting most of the beamline space to a C-band cryogenic photo-injector/linac-based prototype of UC-XFEL. There is space at SAMURAI for 3 linac sections and 3 m  of undulator along with a single compressor, to permit testing an shortened EUV FEL using UC-XFEL/C$^3$ technologies. The layout of SAMURAI that will be realized in two years, with S-band and C-band infrastructure in place, in shown in \ref{fig:SAMURAI}.

The UCLA infrastructure and its goals, as described above, are highly synergistic with  the goals of the C$^3$ project. This synergy is incorporated into proposals for future work, particularly aimed at an NSF STC, to jointly pursue UC-XFELs and advanced accelerators for HEP, such a \CCC.

%\item \textbf{CERF-NM and LEDA building at LANL} 
\subsubsection{CERF-NM and LEDA building at LANL}
\label{sec:LANL}

\begin{figure}[!ht]
\begin{center}
 \includegraphics[trim={0 0cm 0cm 0cm}, clip, width=0.95\hsize]{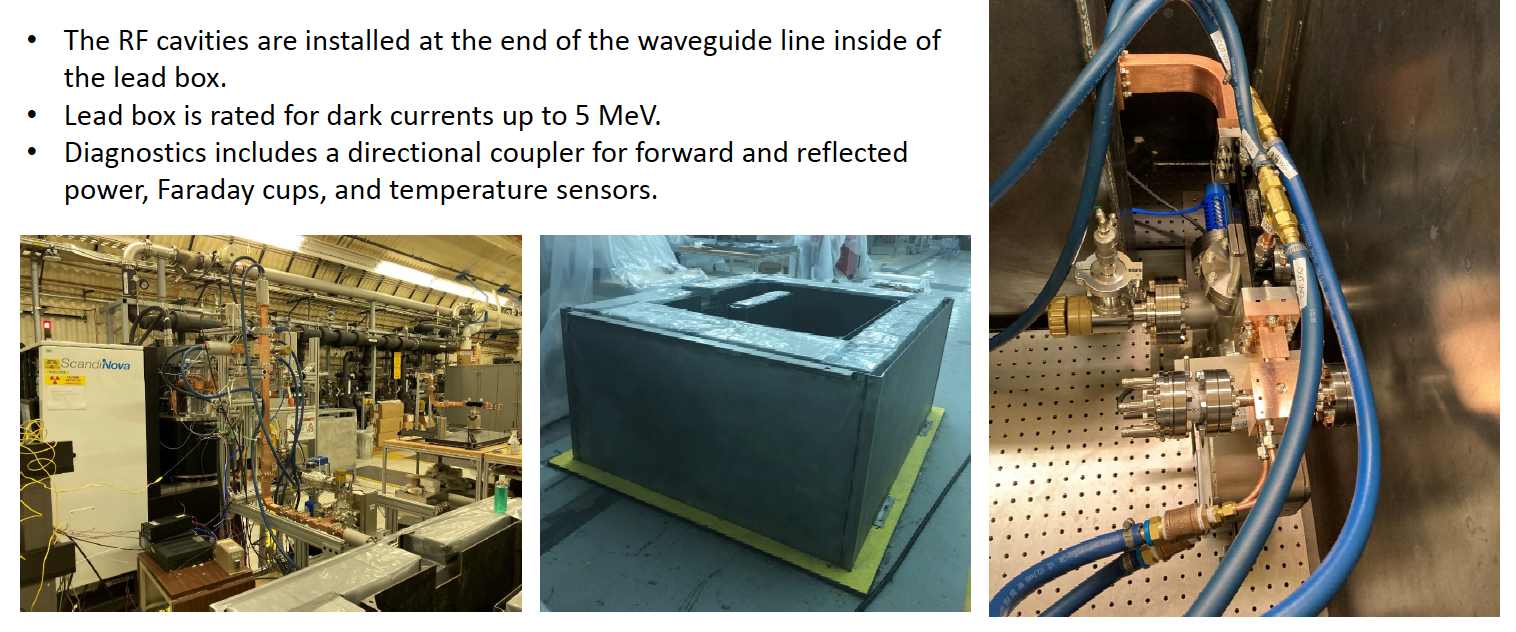} 
\end{center}
\caption{Existing C-band test capabilities at LANL are already playing a crucial role in the development of \CCC technology with rapid testing of structures in excess of 200 MeV/m\cite{schneider2021}.  }
\label{fig:lanltest}
\end{figure}

Los Alamos National Laboratory currently serves as the pioneer in developing the US C-band high gradient testing program. As a part of the program, LANL commissioned the high gradient test stand called C-band Engineering Research Facility – New Mexico (CERF-NM). Photographs of the facility are shown in Figure \ref{fig:lanltest}. The CERF-NM test stand is built around a 50 MW 5.712 GHz Canon klystron. The klystron system produces 50 MW pulses with the pulse length between 300 ns and 1 microsecond, repetition rate up to 200 Hz, and is tunable within the frequency range of 5.707 GHz to 5.717 GHz. The RF power is being output from the klystron in a WR187 rectangular waveguide. The waveguide was conditioned to the full output klystron power of 50 MW, pulse length of 1 microsecond, and repetition rate up to 100 Hz while running the power through the whole length of the line into a water-cooled load. After the conditioning was complete, the breakdowns in the waveguide line became extremely rare. Currently we observe no more than several breakdowns in the waveguide line during the whole day of operation. The WR187 waveguide brings power into a 3 foot by 4 foot lead box designed to protect equipment and operators from X-rays generated in cavities under high gradient testing. The lead box is radiologically certified for dark currents with electron energy up to 5 MeV and average current up to 10 µA. So far, two accelerating cavities were tested for high gradient operation at CERF-NM and surface electric fields in excess of 400 MV/m were demonstrated.  For more details on the CERF-NM test stand and cavity testing see \cite{gorelov2021,simakov2021}.

\begin{figure}[!ht]
\begin{center}
 \includegraphics[trim={0 0cm 0cm 0cm}, clip, width=0.95\hsize]{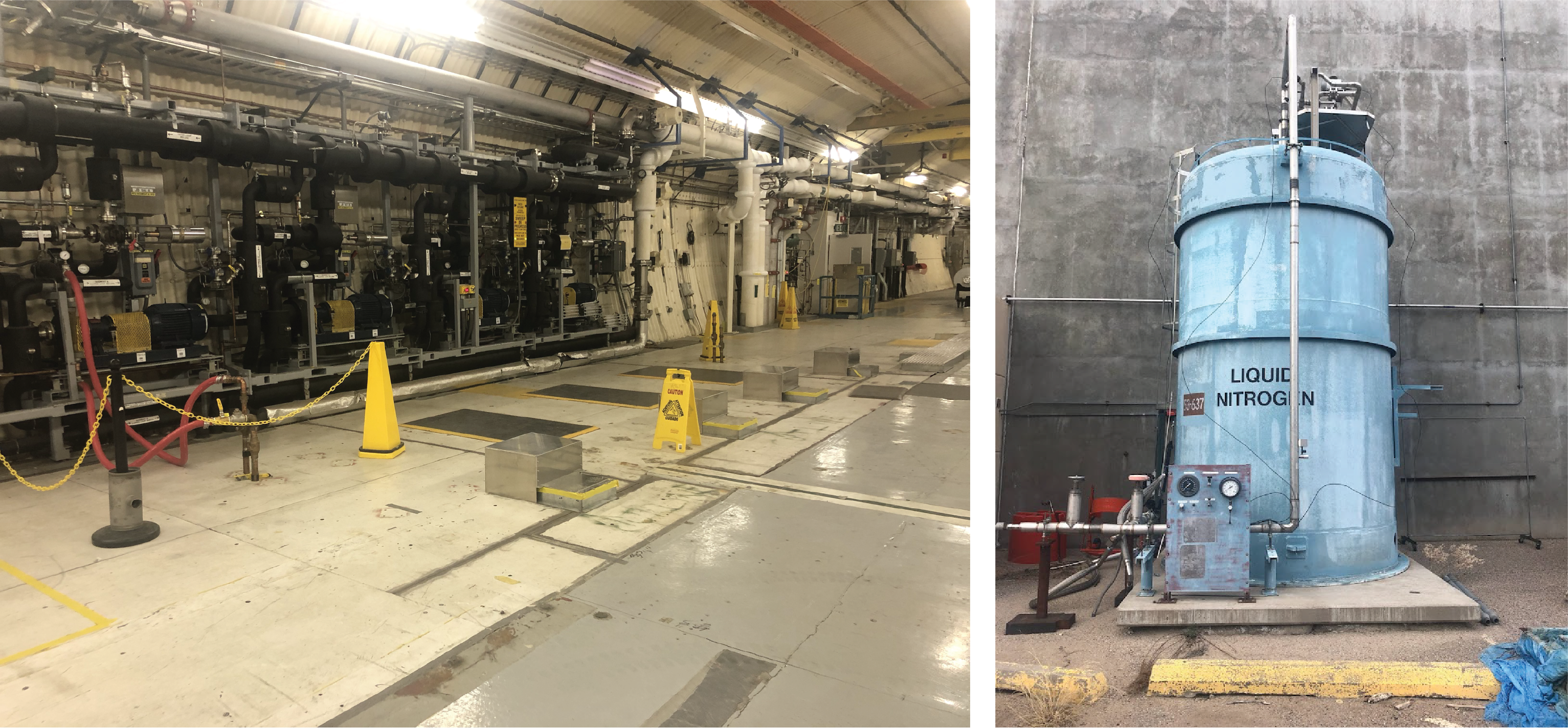} 
\end{center}
\caption{The LEDA building at LANL has a large accelerator tunnel and the cryogen infrastructure needed for the\CCC demonstration facility. }
\label{fig:lanlleda}
\end{figure}

CERF-NM is located in the radiation protection tunnel of the Low Energy Demonstration Accelerator (LEDA) building at Los Alamos Neutron Science Center (LANSCE). The LEDA building and the tunnel offer all necessary infrastructure for locating the proposed \CCC demonstration facility. The dimensions of the tunnel are 470 feet long and 30 feet wide. The tunnel provides adequate radiation protection for up to 10 $\mu$A average electron beam current at the beam energy of 1 GeV. A waveguide basement is located under the tunnel and a laser room with the proper laser infrastructure is at the end of the tunnel. The LEDA building has a large liquid nitrogen tank located next to the building and the required cryogenic infrastructure (see Figure \ref{fig:lanlleda}. The building has multiple options for a control room and rooms for ancillary equipment.

%%%%%%%%%%%%%%%%%%%%%%%%%%%%%%%%%%%%%%%%%%%%%%%%%%%%%%%%%%%%%%%%%%%%%%%%%   
%%%%%%%%%%%%%%%%%%%%%%%%%%%%%%%%%%%%%%%%%%%%%%%%%%%%%%%%%%%%%%

%%%%%%%%%%%%%%%%%%%%%%%%%%%%%%%%%%%%%%%%%%%%%%%%%%%%%%%%%%%%%%%%%%%%%%%%%   
%%%%%%%%%%%%%%%%%%%%%%%%%%%%%%%%%%%%%%%%%%%%%%%%%%%%%%%%%%%%%%

%\end{itemize}

\subsubsection{LEA at Argonne}
\begin{figure}[htbp!]
    \centering
    \includegraphics[width=\textwidth]{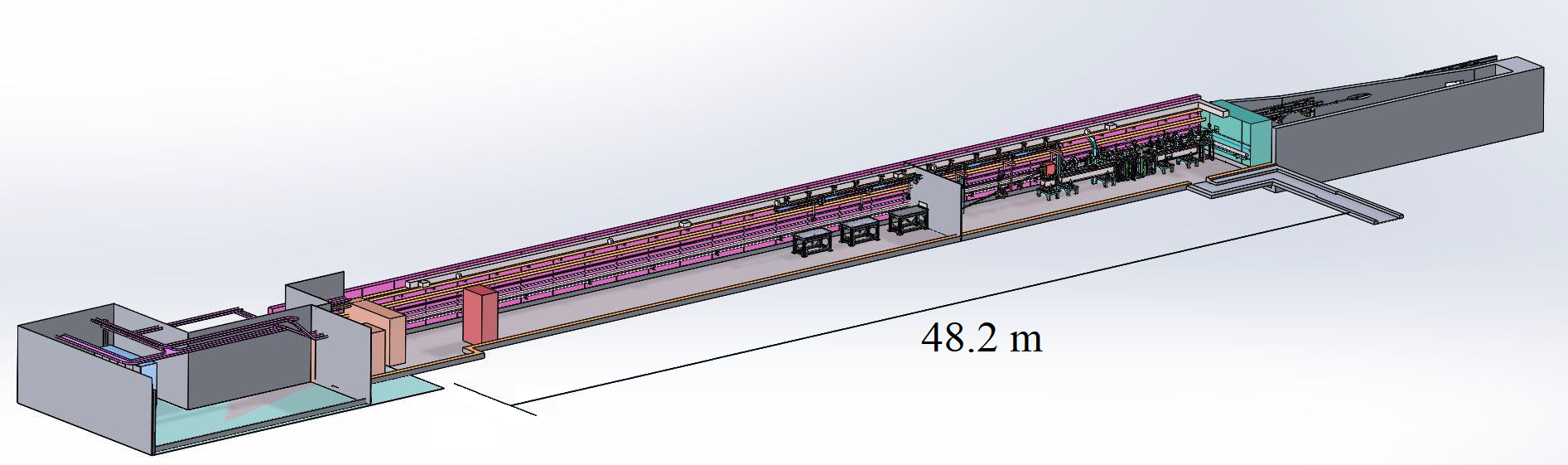}
    \caption{Schematic illustration of the LEA enclosure at Argonne. In this figure, electron beam propagates from right to left. The enclosure length is 48.2\,m. An end-station building is located downstream of the accelerator enclosure.}
    \label{fig:LEA_enclosure}
\end{figure}

The Linac Extension Area (LEA) at the Advanced Photon Source (APS) at Argonne is a flexible accelerator beamline test area and enclosure for demonstrations of new accelerator concepts and technologies \cite{berg_2019}. The LEA enclosure was originally housed the Low-Energy Undulator Test Line (LEUTL) \cite{milton_1997, milton_2001} that made the first demonstration of saturation of the self-amplified spontaneous emission (SASE) process in the visible. The LEA enclosure, together with the APS linac, meet the main requirements of the C$^3$ demonstration system as outlined in Section~\ref{sec:demo_system_basic_req}. The LEA enclosure is illustrated in Fig.~\ref{fig:LEA_enclosure}. LEA is located downstream of the APS linac which can provide electron beam energies up to ${\sim}$500 MeV into LEA. The APS linac also operates with two electron sources: a thermionic RF gun which provides a train of S-band bunches and a low emittance photocathode RF gun.

%%%%%%%%%%%%%%%%%%%%%%%%%%%%%%%%%%%%%%%%%%%%%%%%%%%%%%%%%%%%%%
%%%%%%%%%%%%%%%%%%%%%%%%%%%%%%%%%%%%%%%%%%%%%%%%%%%%%%%%%%%%%%

\begin{figure}[htbp!]
    \centering
    \includegraphics[width=\textwidth]{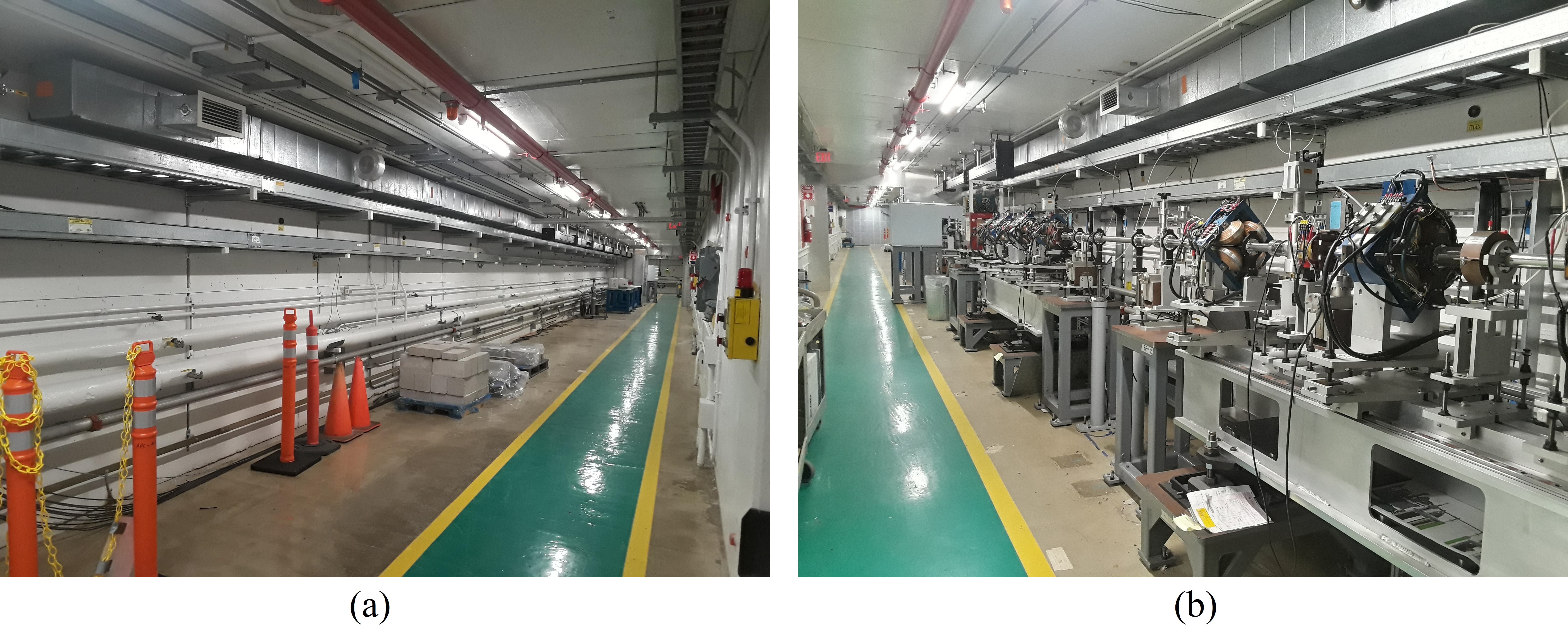}
    \caption{Photographs of the LEA enclosure. (a) LEA enclosure viewed from downstream end. (b) LEA enclosure viewed from upstream end, with the LEA beamline in the foreground.}
    \label{fig:LEA_enclosure_photo}
\end{figure}

The LEA tunnel has more than adequate space for the 3 cryomodule linac (27\,m) demonstration facility. Photographs of the LEA enclosure are illustrated in Fig.~\ref{fig:LEA_enclosure_photo}. The tunnel dimensions are: 3.66\,m (width), 2.74\,m (height), 48.19\,m (length). At the upstream end of the enclosure, the LEA beamline, which has focusing and diagnostics, is presently installed, occupying approximately 17\,m of length. The demonstration facility would not need the full specified 50\,m length since APS already has a photocathode electron gun source. A feature of the present labyrinth configuration is a `movable wall' as part of the downstream labyrinth, allowing large accelerator equipment (${\sim}3$\,m length) to be wheeled in without removing and restacking shielding.

%%%%%%%%%%%%%%%%%%%%%%%%%%%%%%%%%%%%%%%%%%%%%%%%%%%%%%%%%%%%%%
%%%%%%%%%%%%%%%%%%%%%%%%%%%%%%%%%%%%%%%%%%%%%%%%%%%%%%%%%%%%%%

The LEA beamline receives electron beams from the APS linac, which can accelerate electron beams up to 450\,MeV. The APS linac includes as an electron beam source a high-brightness 2856\,MHz photocathode RF gun \cite{sun_2014, sun_2018}. With an appropriate laser, this could allow APS to easily generate the desired C-band compatible pulse format with 5.26\,ns spacing (every 15\textsuperscript{th} S-band bunch), for a full demonstration of C$^3$.

In addition to demonstrating the 3 cryomodule linac, LEA could also be used to test the C$^3$ injector (e.g. positron capture linac and booster linacs described in Section~\ref{sec:accel_structures_main_source_linac}). The LEA enclosure presently has two S-band RF sources. A 30\,MW modulator and klystron and 50\,MW solid state modulator and klystron are located in the LEA endstation (downstream end of beamline). These could be configured to provide RF power to structures within the LEA enclosure, either for experiments with electron beam or structure conditioning. These klystrons could for instance be used to power S-band structures.

The LEA enclosure and beam dump were designed and rated for 700\,MeV beams at 1\,kW of beam power. A new suitably-dimensioned dump could be specified for a 2\,GeV electron beam. An access control and interlock system is installed, and gamma ray and neutron radiation monitors are installed at the upstream and downstream ends of the enclosure. 

A concrete pad is located adjacent to the end-station that could be used to support a liquid nitrogen tank. Truck access to LEA in the APS infield (to refill the nitrogen tank) is via the vehicle tunnel beneath the APS storage ring. The APS receives regular bulk liquid nitrogen deliveries via a service provider at multiple locations around the ring, principally to operate user beamlines.

Utilities including conventional AC power, deionized cooling water and temperature regulation via the building heating, ventilation, and air conditioning (HVAC) system are also installed.

Accelerators at APS are routinely controlled via the Experimental Physics and Industrial Control System (EPICS), and centrally managed from the APS Main Control Room. A laser room has been configured within the LEA end-station.

\subsection{Hardware}

Although the 3 cryomodues contain 24 accelerator sections, we propose to acquire only 18 commercial 50 MW modulator/klystron packages to lower costs. This will permit testing a cryomodule at more than 150 MeV/m to determine where breakdown may set in; to run at 120 MeV/m to demonstrate that breakdown levels at this gradient are small, and to run at 70 MeV/m which we expect for \CCC phase 1. 

The high power RF system for the initial \CCC{} demonstrator will consist of commercially available klystrons and modulators. The HPRF system deployed at LANL for C-band high gradient structure testing is taken as the baseline configuration.\cite{schneider2021}  The LANL test stand utilizes a 5.712 GHz, 50 MW pulsed klystron (Canon E37212) powered by a Scandinova K2-2 modulator. Using this same configuration (or a commercially available equivalent) will support the fastest, lowest risk path to an initial demonstration. Future upgrades to the high power RF system for improved efficiency or reduced fabrication costs, resulting from the direct and parallel R$\&$D initiatives described previously, may be incorporated as they become available.

The \CCC accelerators operate under LN$_2$. The \CCC demonstrator will operate in two modes: The first will be simple operation the linac, with LN$_2$ being fed to one end of the linac with both feed-forward control based on estimated power lost in the copper and feedback based on \LN level sensors. With full operation of the 18 klystrons, this is expected to consume $\sim$25000 liters/day. This will come from a local storage dewar and be replenished by trucks. The minimum local storage tank would be 25,000 liters and would require two truck delivers per day to fill. For continuity of operations, we will pursue  at least double that to have some buffer (weather, traffic, \textit{etc.}). A 50,000 liter storage tank is a fairly common size tank (~13,000 gal). The second mode will demonstrate the full \CCC flow of LN$_2$, $\sim$6 liters/sec,  which will require taking LN$_2$ from the far end at a controlled rate and pumping it back to the storage dewar. Note that for \CCC, each super-sector is fed LN$_2$ from each end. 

The source will be an S Band rf photo-injector that will capable of producing a full \CCC equivalent beam current. The individual bunch charge will be up to 1 nC. This will demonstrate the ability of the structures and sources to maintain the proper gradient over the full train.

\CCC draws heavily on the damping ring and beam delivery/final focus work of ILC~\cite{ilc,aryshev2022international}, NLC~\cite{nlc}, and CLIC~\cite{clic}. Those systems and components are extremely well studied and will not be part of the demonstrator hardware development unless it is necessary.

\section{Acknowledgements}

The work of Mei Bai, Tim Barklow, Martin Breidenbach, Ankur Dhar, Joseph Duris, Auralee Edelen, Claudio Emma, Josef Frisch, Annika Gabriel, Spencer Gessner, John Lewellen, Carsten Hast, Emilio A. Nanni, Cho-Kuen Ng, Anatoly K. Krasnykh, Mohamed A. K. Othman, Marco Oriunno, Dennis Palmer, Michael E. Peskin, Thomas J. Peterson, Emma Snively, Caterina Vernieri, Sami Tantawi, Brandon Weatherford, and Glen White is supported  by Department  of Energy Contract DE-AC02-76SF00515.

Argonne National Laboratory’s work used the resources of the Advanced Photon Source, a U.S. Department of Energy (DOE) Office of Science User Facility operated for the DOE Office of Science by Argonne National Laboratory under Contract No. DE-AC02-06CH11357.

\addcontentsline{toc}{section}{Bibliography}

\bibliographystyle{atlasnote}
\bibliography{bibliography.bib}

\end{document}